\documentclass{aa}

\usepackage{graphicx}
\usepackage{txfonts}
\usepackage{textgreek}
\usepackage{siunitx}
\usepackage[allcolors=blue]{hyperref}
\usepackage{natbib}
\bibpunct{(}{)}{;}{a}{}{,} 
\usepackage[normalem]{ulem}
\usepackage{tabularx}

\newcommand{\crfname}{VIE2023-VG CRF}
\usepackage{color}

\begin{document}

\title{A celestial reference frame derived from observations with the Very Long Baseline Interferometry Global Observing System}
\titlerunning{VGOS CRF 2023}
   \author{H. Kr\'asn\'a \inst{1}
        \and
          C. S. Jacobs \inst{2}
          \and
          M. Schartner \inst{3}
          \and
          P. Charlot \inst{4}
          }

   \institute{ Technische Universit\"at Wien (TU Wien), Department of Geodesy and Geoinformation, Austria\\
              \email{hana.krasna@tuwien.ac.at}
               \and
              Jet Propulsion Laboratory, California Institute of Technology, USA
              \and
              ETH Zurich, Institute of Geodesy and Photogrammetry, Switzerland
              \and
              Laboratoire d’astrophysique de Bordeaux, Univ. Bordeaux, CNRS, France    
             }

   \date{Received 26 August 2024; accepted 07 November 2024}


  \abstract
   {}
   {We computed a celestial reference frame (CRF) from Very Long Baseline Interferometry (VLBI) Global Observing System (VGOS) data after five years of regular observations carried out under the umbrella of the International VLBI Service for Geodesy and Astrometry. In this paper we evaluate its strengths and weaknesses, document the source selection and scheduling strategies for the individual sessions, and investigate the effect of using this new VGOS CRF in the analysis of individual geodetic VLBI sessions where the radio source positions are fixed to their a priori coordinates.
   }
   {We estimated the \crfname{} in a global adjustment of 155 multi-baseline 24-hour VGOS sessions until 2024.0. We carried out several comparisons with  the third version of the International Celestial Reference Frame (ICRF3) adopted by the International Astronomical Union in 2018, and with VIE2023sx CRF which includes VLBI S/X data until 2024.0. 
   Furthermore, we studied the effect of more frequent estimations of tropospheric parameters (30, 10, and 5 min for zenith wet delay) on the estimated CRF in the current VGOS network. We evaluated the \crfname{} in the geodetic analysis of VGOS sessions by computing the baseline lengths and station positions and statistics on the Earth orientation parameters estimated in the single-session analysis where the source positions were fixed to either the \crfname{} or to ICRF3-SX.}
   {The current \crfname{} is built with 1.39 million VGOS group delays and includes 418 radio sources, where 172 sources (41\%) are introduced in only four research and development sessions alone. We show that the \crfname{} has excellent source position precision. The median formal error from the least-squares adjustment is 30~\textmu as for right ascension (scaled by cosine of declination) and 47~\textmu as for declination. In terms of systematic distortions versus ICRF3-SX, the largest terms in the vector spherical harmonics up to the degree and order two, reach in absolute values around 60~\textmu as, caused by correlations between the individual terms. Because of the lack of observations in the southern hemisphere, a constraint for a zero slope in declination difference with respect to ICRF3-SX is imposed in the global adjustment. Therefore, VGOS should prioritize the development of southern stations in order to limit the need for such constraints on the frame.
   Further we show that fixing the a priori CRF to \crfname{} instead of the ICRF3-SX in the single-session analysis improves the weighted root mean square of the baseline length by up to 3~mm, especially for the extremely long baselines (>\num{12 000}~km), with a weighted mean difference between the baseline length time series up to 2~mm. Therefore, in order to meet the ambitious goal of 1~mm accuracy for a terrestrial reference frame set by the Global Geodetic Observing System, the development of a VGOS-only CRF is required for use in the geodetic analysis of the new generation VGOS data.
   }
   {}


   \maketitle
%
\section{Introduction}
The International Very Long Baseline Interferometry (VLBI) Service for Geodesy and Astrometry \citep[IVS;][]{Nothnagel2017} is in transition from its legacy system, which observed radio signals in the S/X-band (2.3/8.4~GHz), to a new broadband VLBI Global Observing System \citep[VGOS;][]{Petrachenko2009, Petrachenko2012}. This was driven by the goals set by the Global Geodetic Observing System (GGOS) of the International Association of Geodesy (IAG), which has envisioned a global terrestrial reference frame (TRF) with an accuracy on a decadal scale of 1~mm and an annual stability of at least 0.1~mm/yr \citep{Blewitt2010, Schuh2001}. A global TRF of such an accuracy is essential, for example for the observation of any subtle geodynamic processes, or glacial and sea ice melting with corresponding sea level changes along the coastlines. VGOS targets the GGOS goals by observing four 512 MHz bands with center frequencies approximately at 3.2~GHz, 5.5~GHz, 6.6~GHz, and 10.4~GHz \citep{Niell2018, Xu2022} with small-diameter (12--13~m) radio telescopes, which have high slew speed in azimuth angle of $\sim12^{\circ}$~s$^{-1}$ and an elevation angle slew speed of $\sim6^{\circ}$~s$^{-1}$ \citep{Petrachenko2009}. This change in the VLBI equipment and observing scheme in these next generation VGOS sessions also has an impact on the determination of the celestial reference frame (CRF), where the source positions are determined from group delays estimated from the linear phase change across the multiple frequency channels \citep{Cappallo2014}.

In this paper we analyze the available multi-baseline 24-hour VGOS sessions. The regular observations have been in place since January 2019, and occur with a weekly to bi-weekly cadence. We provide a detailed description of the source selection for the VGOS sessions (Sect.~\ref{sec:source_list}) and we continue with a precise explanation of the scheduling strategies (Sect.~\ref{sec:scheduling}). We assess the differences in the applied scheduling for the operational sessions (VGOS-OPS) and ten research and development (VGOS-R\&D) sessions, and show the distribution of scheduled scans and observations per source. The analyzed VGOS sessions with the settings applied in the geodetic VLBI analysis software are described in Sect.~\ref{sec_data}. We estimate the celestial reference frame \crfname{} in a global solution of the VGOS sessions, with consistent TRF and Earth orientation parameters (EOP) and provide the CRF characteristics in Sect.~\ref{sec_character}. \\
The main part of the paper (Sect.~\ref{sec_compare}) is devoted to comparisons of the \crfname{} with the third version of the International Celestial Reference Frame \citep[ICRF3;][]{Charlot20}, which is the currently adopted version of the ICRF by the International Astronomical Union (IAU) (Sect.~\ref{sec_compare_icrf3}). We also conduct the comparison with VIE2023sx CRF \citep{Krasna2024_vie2023sx, Krasna2023},\footnote{\url{https://doi.org/10.48436/yt390-wfm94}} which includes five more years of VLBI sessions than ICRF3 (i.e., until December 2023) similar to \crfname{}. In addition, we assess the effects of shorter estimation intervals for the path delays in the neutral atmosphere on the estimated VGOS CRF solutions in the current VGOS network (Sect.~\ref{sec_compare_zwd}).
The evaluations of the \crfname{} in a geodetic analysis (Sect.~\ref{sec_eval_analysis}) are done in terms of baseline length and station position offsets and repeatabilities together with statistics of the estimated five Earth orientation parameters with fixed CRF in the single-session analysis. In Sect.~\ref{sec_conc} we summarize our main findings.

\section{Observed sources} \label{sec:source_list}
For simplicity, we separated the VGOS sessions into three groups which differ in the strategy applied to select the observed sources.
\paragraph{Phase 1:}
VGOS sessions up to VT9318 (2019 November 14) were generated using the same 99 sources selected from a source list that is regularly utilized for sessions within the IVS containing a total of 336 sources.\footnote{\url{https://raw.githubusercontent.com/nvi-inc/sked_catalogs/main/source.cat.geodetic.good}}  
It is important to note that these 99 sources were only considered potential sources during the creation of the schedule. 
Due to scheduling optimization algorithms, source visibility, source brightness, and other factors, not all of them were used in the final schedule, see Sect.~\ref{sec:scheduling}. In fact, the average number of scheduled sources per session during this time was only 67. From VT8067 onwards (2018 March 08), the number of potential sources dropped by one to 98 since 0718+792 was removed from the potential list of sources. 
Surprisingly, two of these sources have never been scheduled (0035$-$252 and 1357+769), while ten others were scheduled in fewer than five sessions. 
\paragraph{Phase 2:}
From VT9329 onwards (2019 November 25), the list of potential sources considered in the VGOS-OPS sessions was extended to cover more sources from the initial IVS source list. 
Since then, the number of considered sources has increased to 284. 
During this time, on average, 77 sources were scheduled per session. 
However, there is a strong imbalance with respect to the number of scans per source per session. 
On average, 10\% of sources were scheduled in fewer than five scans and 25\% in fewer than ten scans.
On the other hand, on average, 10\% of sources were scheduled in more than 45 scans and 25\% in more than 34 scans. 

\begin{figure}
   \centering
   \includegraphics[clip, trim=5cm 8.7cm 4.5cm 8.7cm, width=\hsize]{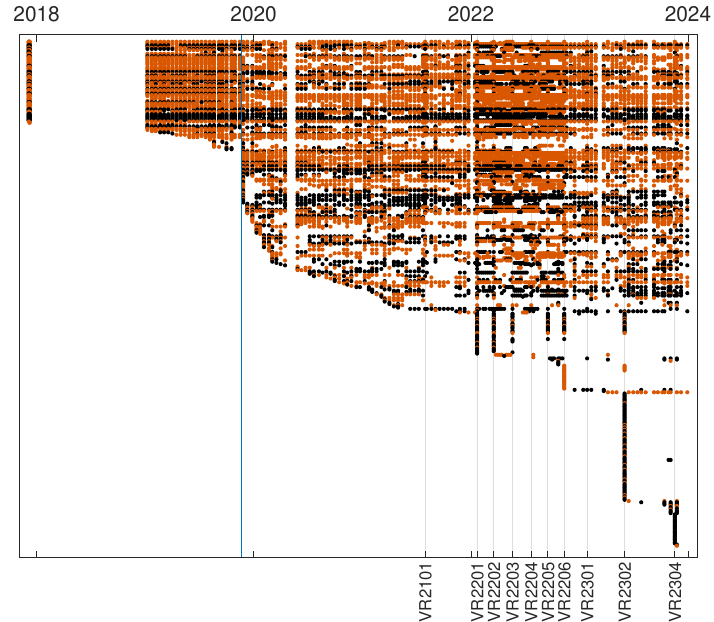}
      \caption{Observed radio sources (y-axis) in \crfname{} per session. The X-axis is sorted chronologically with the VGOS-R\&D sessions marked. The vertical blue line represents the transition between Phase~1 and Phase~2. ICRF3 defining sources are highlighted in orange.}
         \label{Sources_VGOSCRF_orange_defsou}
\end{figure}

\begin{figure*}[]
   \includegraphics[width=\textwidth]{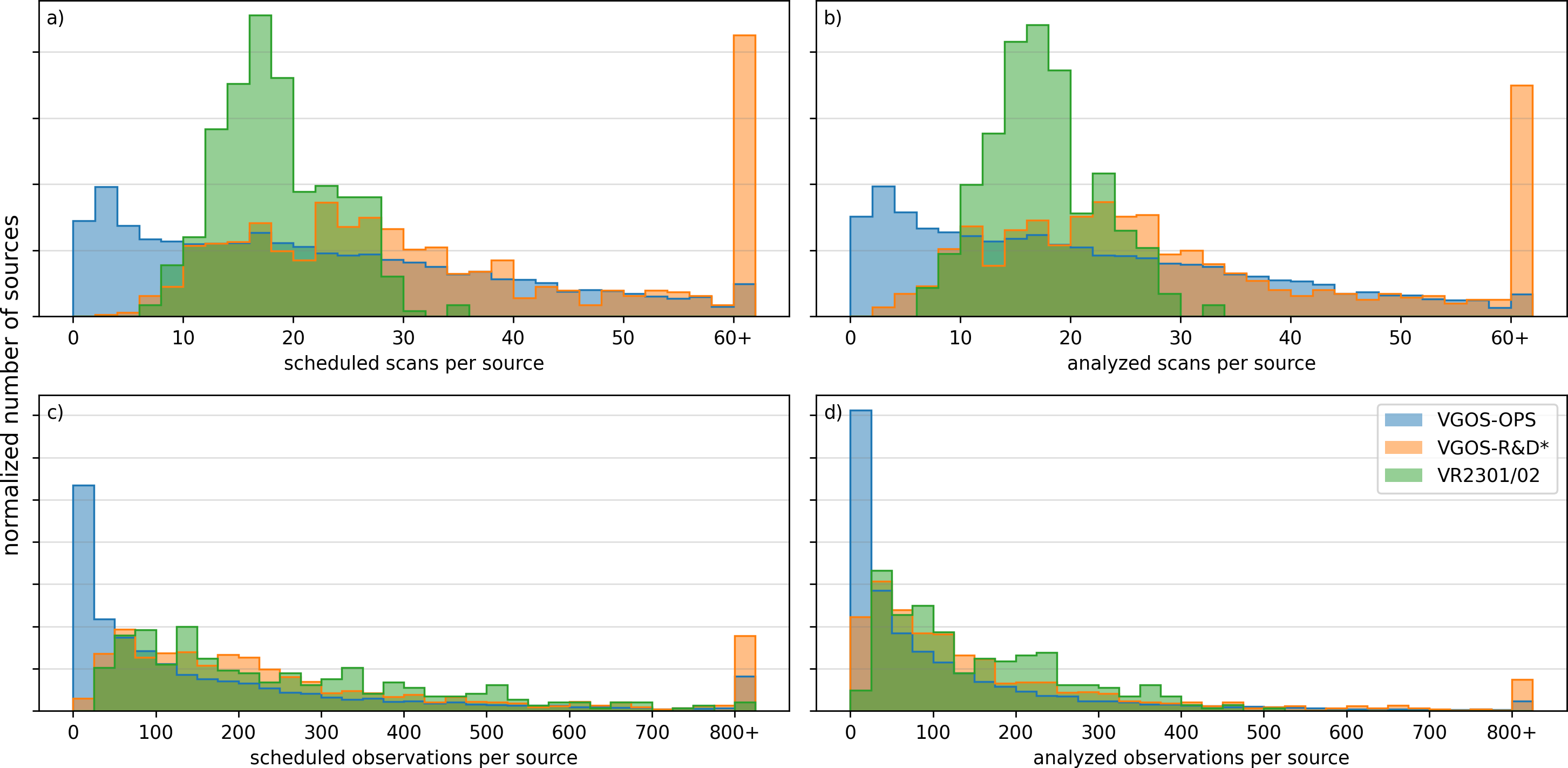} 
      \caption{Histograms with distribution of scans (upper panels) and of observations (lower panels) per source based on the generated schedules (left panels) and on the performed good observations (right panels). The histograms are generated separately from VGOS-OPS (blue), VGOS-R\&D without VR2301 and VR2302 (orange), and only for VR2301 and VR2302  (green). The normalized number of sources means that the total area under each histogram adds up to one. For example, sessions VR2301 and VR2302 observed 114 and 177 sources, respectively, so the original sample size for deriving the VR2301/02 histogram (in green) was 291.}
         \label{sched_analyse_scanobs}
\end{figure*}

\paragraph{VGOS-R\&D:}
The first VGOS-R\&D session VR2101 (2021 July 29, see Fig.~\ref{Sources_VGOSCRF_orange_defsou}), was scheduled using the 99 sources from Phase~1, plus 55 other sources from Phase~2 with low residuals in previous VGOS sessions. 
Thus, the total number of potential sources was 154 out of which 94 sources were scheduled. 

Sources for sessions VR2201--VR2205 (2022 January 20 to September 15) were selected based on the performance in previous VGOS sessions. 
In this context, the performance of a source is defined based on the root mean square of the residuals from observations of this source calculated based on the least-squares adjustment (see Sect.~\ref{sec_data}), as well as based on the success rate of the observations, meaning the fraction of scheduled observations that were finally used in the analysis. 
Furthermore, 62 additional sources were added to the list of potential sources out of which 37 were not in the initial source list and 36 had not yet been scheduled with VGOS. 
These sources were selected by filtering for sources with \SI{400}{mJy} median flux density for baselines > \SI{5000}{km}
at S/X (or C/X in case no S-band flux densities were available) and with the median ratio of X-band flux density over the flux density of the lower band greater than 0.5. 
From the list of new potential sources, 34 were scheduled in VR2201.

VR2206 (2022 November 09) was scheduled with a list of 142 potential sources out of which 101 were selected based on previous VGOS source performance. 
The remaining 41 were selected from the ICRF3 defining source list, among which 
29 were not in the initial source list and 25 had not been scheduled with VGOS. 
From the 41 additional sources, 25 new sources were scheduled in this session. 

Session VR2301 (2023 January 25) and VR2302 (2023 May 31) were scheduled with a dedicated source-centric scheduling approach as discussed in Sect. \ref{sec:scheduling}. 
While VR2301 was scheduled using sources that perform well with VGOS, allowing for a better comparison of the scheduling approaches, VR2302 extended the potential source list significantly. 
Initially, the same source list as utilized in VR2301 was used but extended by another 215 sources that either performed well in the previous sessions or had not yet been observed with VGOS but which were theoretically bright enough to be detected. 
Ultimately, the session included 88 new sources that had not been previously scheduled with VGOS. 

Session VR2303 (2023 August 24) was designed to test sources for bandpass calibration and included another 25 new sources. 
The session was not designed for geodetic or astrometric analysis, thus, it is not representative of the VGOS-R\&D programme. The database was not released by the IVS and therefore is beyond the scope of this paper. 

Finally, VR2304 (2023 November 16) included another 31 potential sources that had so far not been observed, out of which 25 were scheduled in the session. 

In summary, the VGOS-R\&D sessions played a significant role in extending the VGOS source list: 
VR2302 included 88 new sources, VR2201 34, and VR2206 with VR2304 added 25 each.
In total, 172 of the 481 sources scheduled in VGOS were introduced in these four sessions alone, while the potential list of additional sources is still significantly larger.

\section{Scheduling strategy} \label{sec:scheduling}
The VGOS-OPS sessions are scheduled using the "sked" scheduling software \citep{Gipson2010}. 
The scheduling strategy, relevant to the content of the paper, is the following:
in the first step, the software screens the list of potential sources and pre-selects 100 sources that are further considered in the scheduling process (or 99/98 sources for sessions during Phase~1 discussed in Sect.~\ref{sec:source_list}). 
The scheduling is done using a standard geodetic scheduling approach. 
From the 100 pre-selected sources, only a subset is scheduled in any given session. 
Due to the geodetic scheduling approach, there is some severe imbalance with respect to the distribution of scans among sources which is not ideal for deriving a CRF. 
To counter this situation a bit, a total of 26 sessions included a special "astrometric" scheduling option. This option when applied to a source increases its likelihood of being scheduled. 
Ten sources were treated this way in each session for a total of 19 unique sources over the 26 sessions. 
However, it is to be noted that this "astrometric" scheduling option does still not ensure that the listed sources are scheduled in the given session. 
Furthermore, many of the selected "astrometric" sources have already been observed extensively in previous VGOS sessions, such as 0059+581, 0119+115, 0133+476, 0235+164, 0336$-$019, and 0454$-$234, which are among the most-observed VGOS sources. 
Thus, this strategy only partly helps in extending the VGOS source list and cannot be considered an effective strategy to support the generation of a VGOS CRF. 
Finally, it is important to note that all of the potential sources observed within the operational VGOS sessions are from the initial IVS source list, which is only a subset of all sources that can be observed with VGOS. 

The VGOS-R\&D sessions were scheduled using the "VieSched++" scheduling software \citep{Schartner2019}. 
Among the various goals of this research and development program, one goal was to extend the list of observed VGOS sources, as discussed in Sect.~\ref{sec:source_list}. 
As shown in  Fig.~\ref{sched_analyse_scanobs}a, almost all sources were scheduled with a minimum of ten scans per session, allowing for a better determination of the source coordinates. 
Additionally, the scheduling strategy was adjusted several times to ensure a better distribution of scans among sources. 
Most importantly, a new source-based scheduling approach was developed and partly tested in VR2201 and VR2206 before it was further refined and fully utilized in VR2301 and VR2302. 
In essence, it aims to provide a more balanced distribution of scans among sources and time. The green histogram in the upper left plot in Fig.~\ref{sched_analyse_scanobs}a shows that the majority of sources in VR2301 and VR2302 were scheduled with in between ten to 30 scans which already provides a good basis for a CRF determination, while the distribution is very inhomogeneous in the standard geodetic VGOS-OPS or VGOS-R\&D sessions. 
The scheduling strategy works by first defining a target number of scans per source, followed by calculating the total up-time of each source based on the given network and a minimum number of stations per scan. 
Based on these metrics, the optimal repeat time between two scans to the same source is calculated.
With this repeat time, the target number of scans can be reached while ensuring a good distribution of scans over time. 
A further benefit from distributing the scans equally over time is that this way, it also results in a good uv-coverage of the scheduled sources for increased imaging potential.
Furthermore, it is likely that a good distribution of scans among sources and in time also improves the geodetic performance of the session since unmodelled systematic effects might be averaged out more equally. 
More detailed information on the scheduling mode can be found in \citet{Schartner2023b}. 

Due to the imbalanced VGOS network geometry (see Sect.~\ref{sec_data}) and the complex nature of VLBI scheduling, it is not possible, nor required, to perfectly balance the distribution of scans among sources. 
Still, the described scheduling strategy used in VR2301 and VR2302 significantly improved the distribution:
The 0.10, 0.25, 0.75, and 0.90 quantiles of scans per source are 12, 15, 21, and 25 respectively, which is more balanced compared to the VGOS-OPS session with quantiles of 5, 10, 34 and 45, respectively, as discussed in Sect.~\ref{sec:source_list}. 
The green histogram in Fig.~\ref{sched_analyse_scanobs}a further shows that the distribution is also improved compared to that from the eight VGOS-R\&D* sessions (i.e., all available VGOS-R\&D except VR2301 and VR2302) which include a high number of sources with more than 60 scans. Additionally, significantly more sources can be scheduled per session, which in turn calls for enlarging the pool of suitable sources for VGOS. 
Session VR2301 observed a total of 114 sources while session VR2302 observed 177 sources. On the other hand, the number is on average 77 for the VGOS-OPS sessions. 
The difference in the number of sources between VR2301 and VR2302 can be explained by other scheduling settings. While VR2301 followed the operational VGOS approach of observing each scan for 30 seconds, VR2302 utilized an observing time based on the signal-to-noise ratio, resulting in an increase in the average number of scans per station per hour from 35 (VR2301) to 60 (VR2302) which is reflected in the number of observed sources. 
The histogram of the number of observations of the sources for the same three group of sessions (Fig.~\ref{sched_analyse_scanobs}c) highlights the distribution of observations from VGOS-R\&D* and from VR2301 and VR2302. There are no sources with fewer than 25 scheduled observations in the VR2301 and VR2302 sessions and very few in the VGOS-R\&D* sessions. On the other hand, there is a high number of sources, which were scheduled for more than 800 observations in VGOS-R\&D*. The right side plots in Fig.~\ref{sched_analyse_scanobs} give the discussed histograms for the good observations, which were obtained during the observing sessions and successfully processed with VLBI analysis software. These form the basis of the current \crfname{} \citep{Krasna2024}.\footnote{\url{https://doi.org/10.48436/m6j36-s1082}}

In conclusion, based on the statistics displayed in Fig.~\ref{sched_analyse_scanobs}, the scheduling strategy utilized in VR2301 and VR2302 is beneficial for generating a CRF since it allows the observation of more sources with a better distribution of scans among them. 

\section{Data and analysis settings}
\label{sec_data}

\begin{table*}
\caption{Parameterization of the \crfname{} solution, based on piece-wise linear offsets (PWLOs).}
\label{tab:solparam}       
\begin{tabular}{lrr}
\hline\hline\noalign{\smallskip}
 Parameterization options&  time interval between PWLOs & relative constraints between offsets\\
\noalign{\smallskip}\hline
clocks & 1 h &43 ps (13~mm) \\
zenith wet delay & 30 min &50~ps (15~mm)\\
tropospheric gradients &  3 h & 0.5~mm\\
ERP  &  24 h & 1~mas \\
celestial pole offsets & 24 h &0.1~\textmu as \\
\hline\noalign{\smallskip}
\end{tabular}
\end{table*}

\begin{figure}[h]
   \centering
   \includegraphics[width=0.5\textwidth]{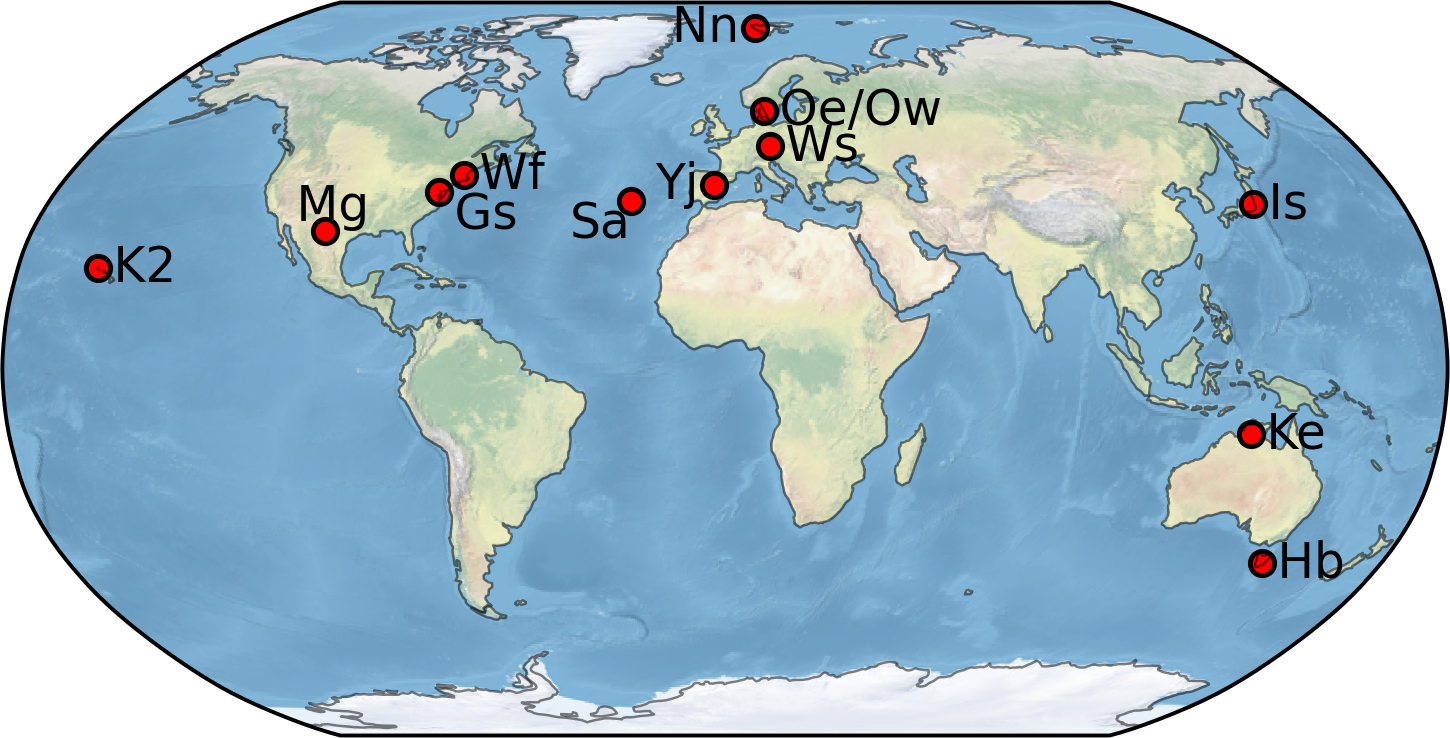}
      \caption{Telescopes in the VGOS network: GGAO12M (Gs), HOBART12 (Hb), ISHIOKA (Is), KATH12M (Ke), KOKEE12M (K2), MACGO12M (Mg), NYALE13N (Nn), ONSA13NE (Oe), ONSA13SW (Ow), RAEGSMAR (Sa), RAEGYEB (Yj), WESTFORD (Wf), WETTZ13S (Ws).}
         \label{fig:network}
\end{figure}

The \crfname{} was computed from all 24-hour multi-baseline VGOS sessions introduced in Sects.~\ref{sec:source_list} and ~\ref{sec:scheduling}. The first five sessions were observed in December 2017 and regular bi-weekly VGOS sessions started in January 2019. Our dataset ends in December 2023 which makes 155 VGOS sessions in total that are provided through the IVS. 
The VGOS telescope network used within these 155 sessions included a total of 13 stations, as depicted in Fig.~\ref{fig:network}. 

\begin{figure}
   \centering
   \includegraphics[clip, trim=5cm 8.5cm 3cm 9cm, width=\hsize]{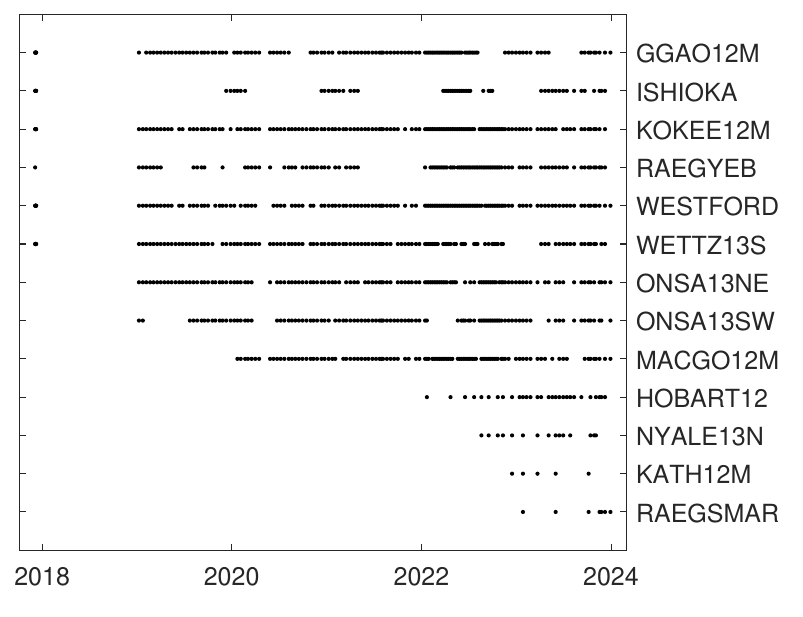}
      \caption{Participating VGOS telescopes in the \crfname{} solution per session.}
         \label{Ant_VGOSCRF}
\end{figure}

Our analysis starts with the VLBI group delays, the fundamental observables of geodetic and global astrometric VLBI, which are retrieved after correlation, fringe fitting and pre-processing from IVS databases in vgosDB format \citep{Gipson2021} via the IVS Data Centers. The analysis is done with Vienna VLBI and Satellite Software  \citep[VieVS;][]{Boehm18} following the standard a priori models and setup as described in \citet{Krasna2023}. Table~\ref{tab:solparam} lists the basic parameterization options chosen for the solution. The weighting of the individual measurements includes an elevation dependent term to account for tropospheric noise \citep{Gipson08} with a special treatment for station HOBART12 (Hb). In 2023, a spurious pattern appeared in Hb phase calibration signal (Lucia McCallum, personal communication) which makes the residuals for Hb baselines noisy. Since Hb is a crucial station in this dataset regarding observation of southern sources (cf. Fig.~\ref{fig:network}), we kept the Hb observations in the adjustment, but we down-weighted them with an additional variance of (5~cm)$^2$ in order to avoid the propagation of the noise via datum definition to other baselines. The a priori positions of the antennas and radio sources were taken from ITRF2020 \citep{Altamimi23} and ICRF3-SX \citep{Charlot20}, respectively. The new NYALE13N (Nn) telescope started its observations in 2022 meaning that its position is not included in ITRF2020. Therefore, we took its a priori coordinates from the recent VLBI terrestrial frame VIE2023 TRF \citep{Krasna2024_vie2023}.\footnote{\url{https://doi.org/10.48436/76404-a3492}}

The datum definition of the global solution was done with no-net-translation (NNT) and no-net-rotation (NNR) condition with respect to the ITRF2020 positions of all included VLBI telescopes except for ISHIOKA (Is) and RAEGSMAR (Sa). The linear velocities of all telescopes were fixed to the a priori values and only the offset in position was estimated (see Fig.~\ref{Ant_VGOSCRF} for participation of the VGOS telescopes in the \crfname{} solution sorted chronologically on the x-axis).
The usual way of applying the celestial datum definition is the NNR condition which forces the sum of cross products between the corrections to the angular coordinates and the source positions components themselves over all included sources to be a zero vector \citep[see][]{Jacobs2010}. The NNR condition was applied on 71 ICRF3 defining sources with more than \num{1000} observations in \crfname{}. In addition, a constraint for a zero slope in declination estimates versus declination was applied. The reason is the lack of observations in the southern hemisphere for defining radio sources---only 22 datum sources are located in the southern hemisphere with four of them having declination lower than~$-20\degr$.

\section{Characteristics of \crfname{}}
\label{sec_character}

\begin{table*}
\caption{Basic characteristics of \crfname{} provided for different groups of sources.}
\label{tab:crfcharact}       
\begin{tabular}{lrrr}
\hline
\hline\noalign{\smallskip}
 Parameter&  all sources& ICRF3 defining& \crfname{} defining\\
\noalign{\smallskip}\hline
no. of sources & 418 & 178 & 71\\
raw formal error $\alpha^*$ & 30~\textmu as & 23~\textmu as & 7~\textmu as\\
raw formal error $\delta$ & 47~\textmu as&  41~\textmu as& 17~\textmu as\\
correlation coeff. between $\alpha$ and $\delta$  &  0.34&  0.39 &  0.35 \\
semi-major axis of error ellipse & 50~\textmu as & 44~\textmu as & 18~\textmu as \\
semi-minor axis of error ellipse & 27~\textmu as & 21~\textmu as & 6~\textmu as \\
\hline\noalign{\smallskip}
\end{tabular}
\tablefoot{Statistics are computed for the entire catalog of all 418 sources, for all observed ICRF3 defining sources, and for ICRF3 defining sources with more than 1000 observations in \crfname{} which were used for datum definition. The reported stochastic characteristics are median values.}
\end{table*}

\begin{figure}[t!]
   \centering
   \includegraphics[clip, trim=2cm 10cm 1.5cm 9cm, width=\hsize]{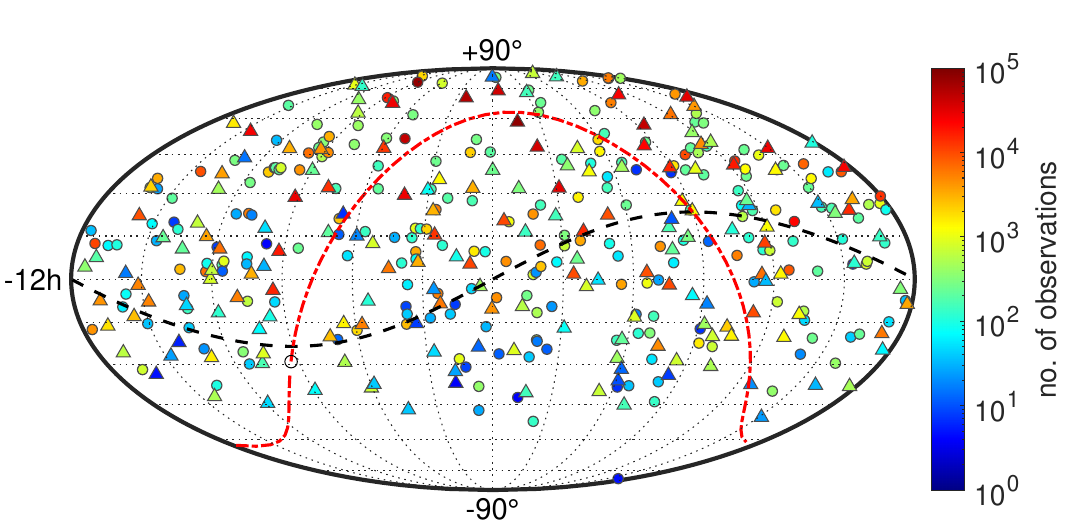}
      \caption{Distribution of all sources with the  number of observations in \crfname\ color-coded (see color bar at right){}. Defining ICRF3 sources are depicted as triangles, non-defining as circles.}
         \label{noObs_aitoff}
\end{figure}

The estimated \crfname{} is built with $1.39\cdot10^6$ observations collected after five years of regular observations of the VGOS network. The distribution of individual sources is depicted in Fig.~\ref{noObs_aitoff} where the triangles represent ICRF3 defining sources. There are 418 radio sources in total, 19 of them with fewer  than 10 observations, 93 sources with between 10 and 100 observations and the 37 most observed sources with more than $10^4$ observations. The number of observations of southern sources is $9\cdot10^4$ which makes up only 6\% of the total number of observations. 
Regarding the number of sessions, 141 sources were observed in one session only, and 31 sources in two sessions only. On the other hand, 91 sources were included in more than 50 sessions (see Fig.~\ref{Sources_VGOSCRF_orange_defsou}).

\begin{figure}
   \centering
   \includegraphics[clip, trim=2cm 8.5cm 2cm 9cm, width=\hsize]{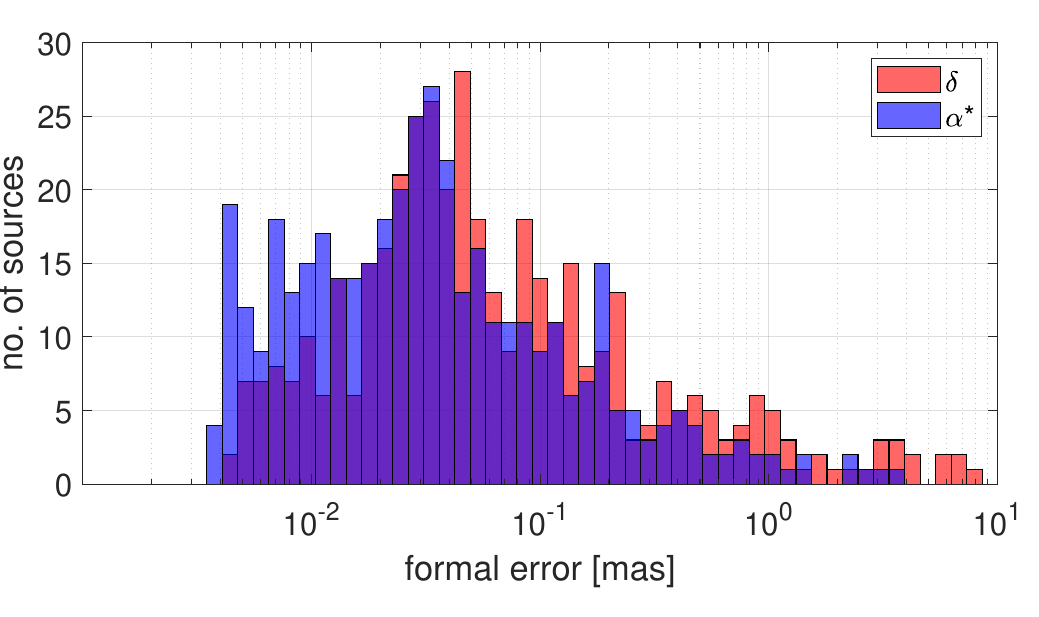}
      \caption{Histogram of raw formal errors of source positions from the \crfname{} solution.}
         \label{CRFfe_histogram}
\end{figure}

\begin{figure*}
   \centering
   \includegraphics[clip, trim=2cm 10cm 1.5cm 9cm, width=0.5\hsize]{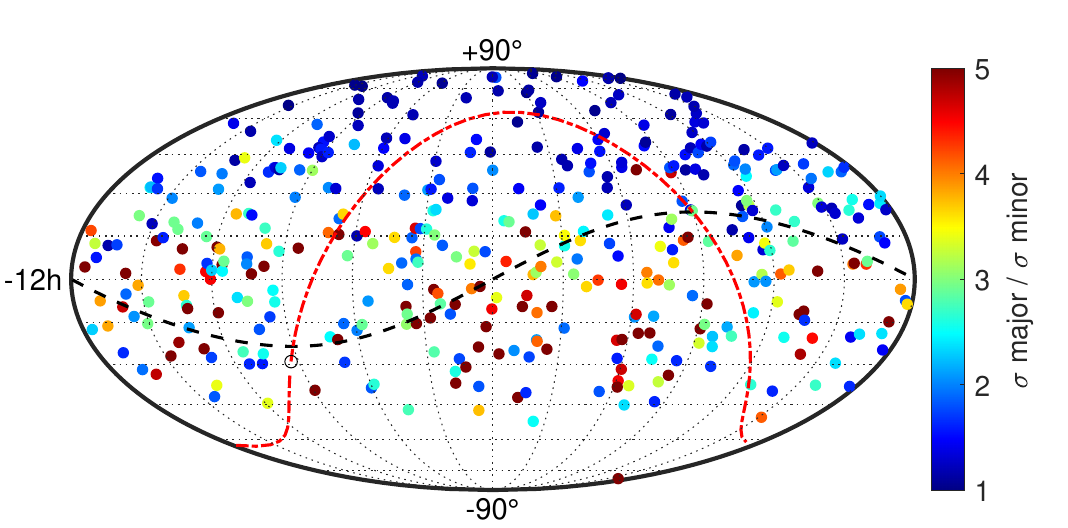}\includegraphics[clip, trim=2cm 10cm 1.5cm 9cm, width=0.5\hsize]{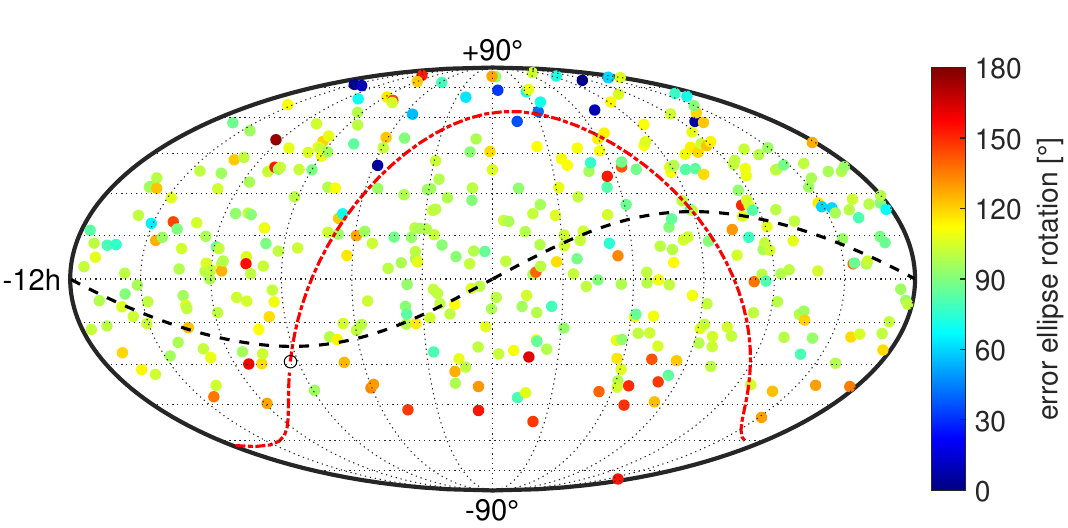}
   \includegraphics[clip, trim=3.8cm 13.8cm 4.2cm 9cm, width=0.5\hsize]{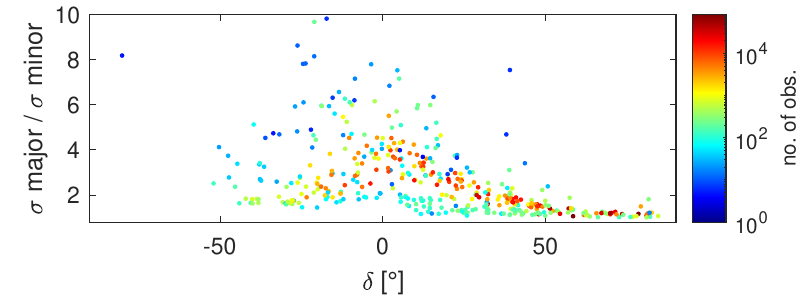}\includegraphics[clip, trim=3.8cm 13.8cm 4.2cm 9cm, width=0.5\hsize]{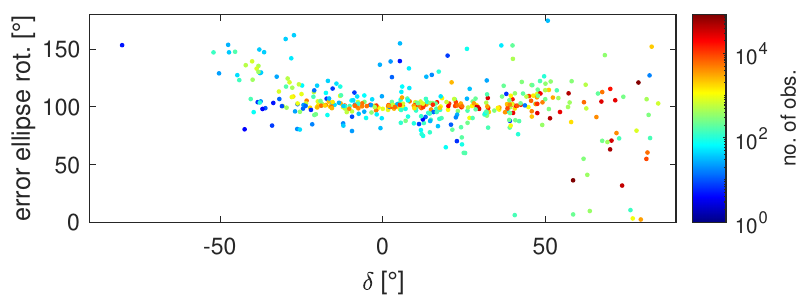}
      \caption{Characteristics of error ellipses in \crfname{}. The panels on the left depict the ratio of semi-major to semi-minor axes for all sources. In the panels on the right the direction of the semi-major axes is shown, counting counterclockwise from the east. The lower panels show the error ellipse characteristics with respect to declination of the sources with a color-coded number of observations.}
         \label{CRFellipse}
\end{figure*}

The raw formal errors, derived directly from the global adjustment, are depicted in Fig.~\ref{CRFfe_histogram}. The median formal errors (Table~\ref{tab:crfcharact}) over all sources are 30~\textmu as for $\alpha^*$ ($\alpha^*$ is right ascension scaled by the cosine of declination ($\delta$) of the source, that is, $\alpha^* = \alpha \cdot \cos{\delta}$) and 47~\textmu as for $\delta$. Median formal errors for ICRF3 defining sources with more than \num{1000} observations in \crfname{} (used for datum definition) are 7~\textmu as for $\alpha^*$ and 17~\textmu as for $\delta$. A careful investigation of the accuracy of estimated source positions from the VGOS sessions needs to be done in order to report realistic uncertainties instead of the raw formal errors and will be a topic for our follow-on research. Note that in the case of the ICRF3-SX, the raw formal errors were scaled by 1.5 and a noise floor of 30~\textmu as was added in quadrature to each coordinate uncertainty~\citep{Charlot20}. The ratio of the semi-axes of the error ellipse for each source is depicted in the left panels in Fig.~\ref{CRFellipse}. Considering sources that were observed repeatedly (more than one thousand times), the ratio of semi-axes shows steady elongation starting with one for the northern declinations increasing to four for sources with zero declination. For poorly observed sources, the large ratio of error ellipse semi-axes shows that the uncertainty in one direction is more than five times worse compared to the uncertainty in the perpendicular direction. The right panels in Fig.~\ref{CRFellipse} depict that the semi-major axes are mostly in the north-south directions (rotation about $90\degr$ counting from the east) with the median value $101.2\degr$ meaning that the estimates in declination are weaker than in right ascension. The way to improve the declination estimates of the sources in the south is to enlarge the VGOS observing network with more telescopes located in the southern latitudes. The addition of stations at Hartebeesthoek, South Africa, and La Plata in South America would be especially helpful \citep{Hase2014, Mayer2014, Gomez2024}.

\section{Comparison and evaluation}
\label{sec_compare}

\subsection{Comparison of \crfname{} to ICRF3-SX}
\label{sec_compare_icrf3}
\begin{figure}
   \centering
   \includegraphics[clip, trim=3.9cm 8.4cm 4cm 9cm, width=\hsize]{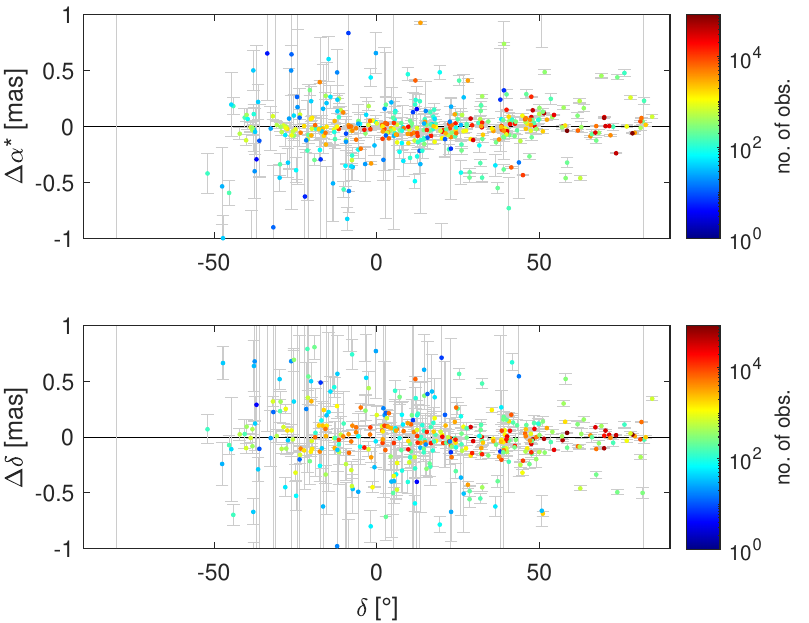}
      \caption{Differences in \crfname{} source position with respect to ICRF3-SX with color-coded number of observations in \crfname{}. The wrms of $\Delta\alpha*$ and $\Delta\delta$ are 80~\textmu as and 100~\textmu as, respectively, computed over sources with normalized angular separation lower than~5.}
         \label{dRADEwrtICRF3sx}
\end{figure}

We carried out a comparison to the IAU-adopted CRF estimated from S/X observations, which is currently the ICRF3-SX. The ICRF3 was computed in 2018 with S/X observations collected from August 1979 until March 2018.
The differences of estimated coordinates in \crfname{} with respect to ICRF3-SX are shown in Fig.~\ref{dRADEwrtICRF3sx} where the colors represent the number of observations to each source in the analyzed VGOS dataset.

\begin{table*}
\caption{Sources with a normalized angular separation $D_n$ between \crfname{} and ICRF3-SX larger than 15.}
\label{tab:souout}       
\begin{tabular}{l l r r r r r r r r r r r r}
\hline\noalign{\smallskip}\hline\noalign{\smallskip}
IERS  & IVS & $\Delta\alpha^*$  & $\Delta\delta$ & 
$X_{\alpha^*}$ & $X_{\delta}$ & $C$ &  $D_n$&
 no. of  & no. of & SI &  Date of SI\\
name  & name  &  [\textmu as] & [\textmu as] &  & & & & sess.  &  obs. & X-band &    from BVID\\
\noalign{\smallskip}\hline
0119+115 & - &        412 $\pm$     31 &        521 $\pm$     37&         13 &         14  & -0.065 &   19.9 &     93 &   6986 &  3.5 & 2018-02-20\\ 
0212+735 & - &        439 $\pm$     38 &       -495 $\pm$     39&         12 &        -13  & -0.074 &   16.5&      1 &    261 &  3.4 & 2016-11-30\\ 
0229+131 & - &        927 $\pm$     32 &        -51 $\pm$     43&         29 &         -1  & -0.132 &   28.9&      52 &   3443 &  3.4 & 2015-03-25\\ 
0355+508 & NRAO150  & -43 $\pm$     39 &       -688 $\pm$     42&         -1 &        -16  & -0.009 &   16.2&     40 &   1992 &  2.8 & 1997-01-10\\ 
0642+449 & - &       -435 $\pm$     31 &       -212 $\pm$     31&        -14 &         -7  & -0.040 &   15.9&      38 &  10073 &  3.1 & 2021-11-24\\ 
0723$-$008 & - &      -2252 $\pm$    147 &       3867 $\pm$    145&        -15 &         27  & -0.580 &   26.7&       1 &     28 &  3.3 & 1995-04-12\\
0738+313 & - &         32 $\pm$     46 &       1796 $\pm$     60&          1 &         30  & -0.039 &   30.2 &      1 &    141 &  4.1 & 2018-08-21\\
0923+392 & 4C39.25  & 739 $\pm$     34 &        -97 $\pm$     44&         22 &         -2  & -0.116 &   21.6 &       2 &    418 &  3.3 & 2013-07-24\\
1548+056 & - &         62 $\pm$     66 &      -1563 $\pm$     96&          1 &        -16  & -0.238 &   16.6&      1 &     87 &  4.2 & 2013-04-10\\
1745+624 & - &       -461 $\pm$     37 &       -461 $\pm$     38&        -13 &        -12  &  0.058 &   17.1&     13 &    608 &  3.4 & 2021-11-24\\ 
\noalign{\smallskip}\hline
\end{tabular}
\tablefoot{The $X_{\alpha^*}$ and $X_{\delta}$ stand for normalized coordinate differences, $C$ is the correlation coefficient of combined
errors, SI stands for source structure index \citep{Fey1997}. The number of sessions and observations refer to \crfname{}.}
\end{table*}

\begin{table*}
\caption{Weighted standard deviation (wstd, Cols. 2--5) and weighted Allan deviation (wadev, Cols. 6--9) of position estimates in $\alpha^*$ and $\delta$, with the difference of weighted mean ($\Delta$w.mean, VGOS minus SX sessions) in $\alpha^*$ and $\delta$ (Cols. 10--11).}
\label{tab:souwstd}       
\begin{tabular}{lrrrrrrrrrr}
\hline
\hline\noalign{\smallskip}
 &  \multicolumn{2}{c}{wstd$_{\text{SX}}$} &\multicolumn{2}{c}{wstd$_{\text{VGOS}}$}
 &  \multicolumn{2}{c}{wadev$_{\text{SX}}$} &\multicolumn{2}{c}{wadev$_{\text{VGOS}}$}&\multicolumn{2}{c}{$\Delta$w.mean} \\
 source& ${\alpha^*}$& $\delta$&$\alpha^*$& $\delta$& ${\alpha^*}$& $\delta$&$\alpha^*$& $\delta$&$\alpha^*$& $\delta$ \\
\noalign{\smallskip}\hline
0119+115 & 221 & 560 & 366 & 757 & 198 & 404 & 221 & 506 &  185 & 25 \\
0229+131 & 325 & 200 & 264 & 435 & 260 & 206 & 268 & 411 &  616 & -18\\
NRAO150  & 285 & 593 & 96 & 716  & 499 & 489 & 101 & 572 &  -50 & -648\\
0642+449 & 152 & 115 & 430  & 82 & 133 & 116 & 182 & 68  & -249 & -89 \\
\hline\noalign{\smallskip}
\end{tabular}
\tablefoot{The wstd, wadev, and the weighted mean are computed for four outlier sources from available S/X and VGOS sessions (including more than 20 observations to the given source) between 2019.0 and 2024.0.}
\end{table*}

Different methods exist for outlier detection as was discussed by \citet{Lambert2023} or \cite{Malkin2024}, among others. They show that the several possible methods detect a similar group of radio sources. We compute the normalized angular separation $D_n$ following \citet{Mignard2016} from normalized coordinate differences $X_{\alpha^*}$, $X_{\delta}$:
\begin{equation}
X_{\alpha^*} = \frac{\Delta\alpha^*}{\sigma_{\Delta\alpha^*}} = \frac{\Delta\alpha^*}{
\sqrt{\sigma^2_{\alpha^*,VG} + \sigma^2_{\alpha^*,SX}}
},
\end{equation}
\begin{equation}
X_{\delta} = \frac{\Delta\delta}{\sigma_{\Delta\delta}} = \frac{\Delta\delta}{
\sqrt{\sigma^2_{\delta,VG} + \sigma^2_{\delta,SX}}
}
\end{equation}
and taking into account the correlation coefficient $C$ of the combined formal errors $\sigma$ reported in the CRF catalogs: 
\begin{equation}
C = \frac{\sigma_{\alpha^*,VG} \cdot \sigma_{\delta,VG} \cdot C_{VG} + \sigma_{\alpha^*,SX} \cdot \sigma_{\delta,SX} \cdot C_{SX}}
{\sqrt{\left( \sigma_{\alpha^*,VG}^2 +  \sigma_{\alpha^*,SX}^2 \right) \left( \sigma_{\delta,VG}^2 +  \sigma_{\delta,SX}^2 \right) }},
\end{equation}
as
\begin{equation}
D_n^2 = \begin{bmatrix} X_{\alpha^*} &  X_{\delta} \end{bmatrix} 
\begin{bmatrix}
  1 & C\\ 
  C & 1
\end{bmatrix}^{-1}
\begin{bmatrix} X_{\alpha^*} \\  X_{\delta} \end{bmatrix} .
\end{equation}

We identified ten radio sources with $D_n$ between \crfname{} and ICRF3-SX greater than 15, and 85 radio sources with $D_n$ greater than 5. The weighted root mean square (wrms) of $\Delta\alpha*$ and $\Delta\delta$ are 142~\textmu as and 166~\textmu as, respectively, computed over sources with $D_n<15$, and 80~\textmu as and 100~\textmu as for sources with $D_n<5$.

The ten radio sources with $D_n>15$ are listed in Table~\ref{tab:souout}. In the last two columns, we included the source structure index \citep[SI;][]{Fey1997} for X-band observations of the sources provided in the Bordeaux VLBI Image Database \citep[BVID,\footnote{\url{ https://bvid.astrophy.u-bordeaux.fr}}][]{Collioud2019} and the date of the S/X session used for the image. We picked up the most recent X-band image provided in the BVID which had a proper resolution from long intercontinental baselines. All of these ten sources have the SI around 3 and 4, which indicates spatially extended sources. Five of them were observed in fewer than three VGOS sessions. 

\begin{figure*}[htb]
   \includegraphics[clip,  trim=7cm 5cm 7cm 5cm,width=0.49\hsize]{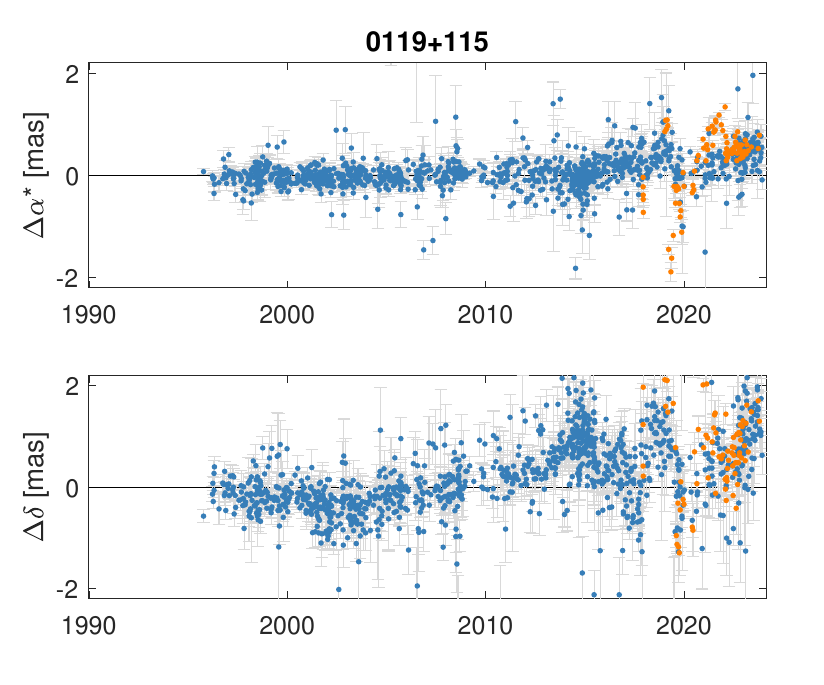} 
   \includegraphics[clip,  trim=7cm 5cm 7cm 5cm,width=0.49\hsize]{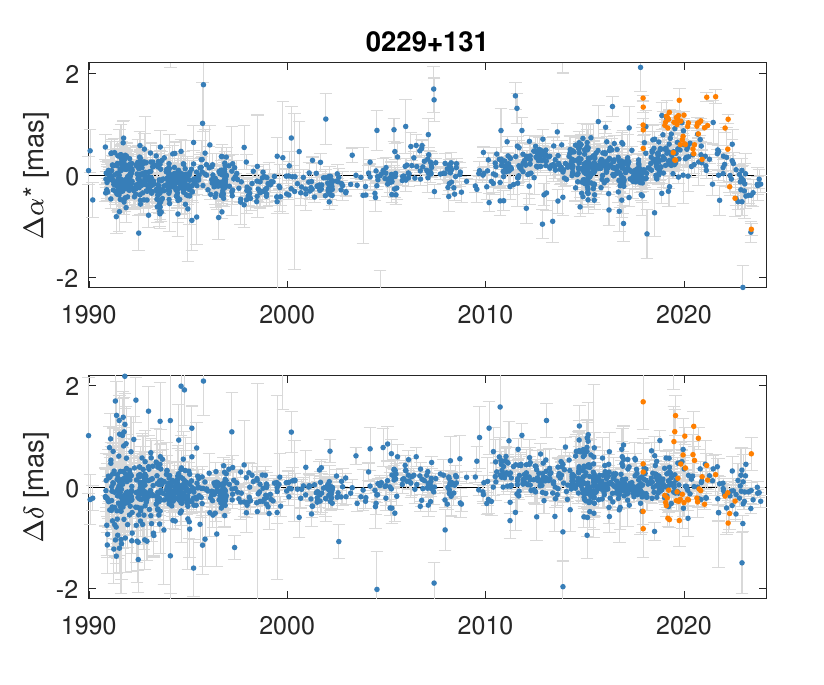}\\
      \includegraphics[clip,  trim=7cm 5cm 7cm 5cm,width=0.49\hsize]{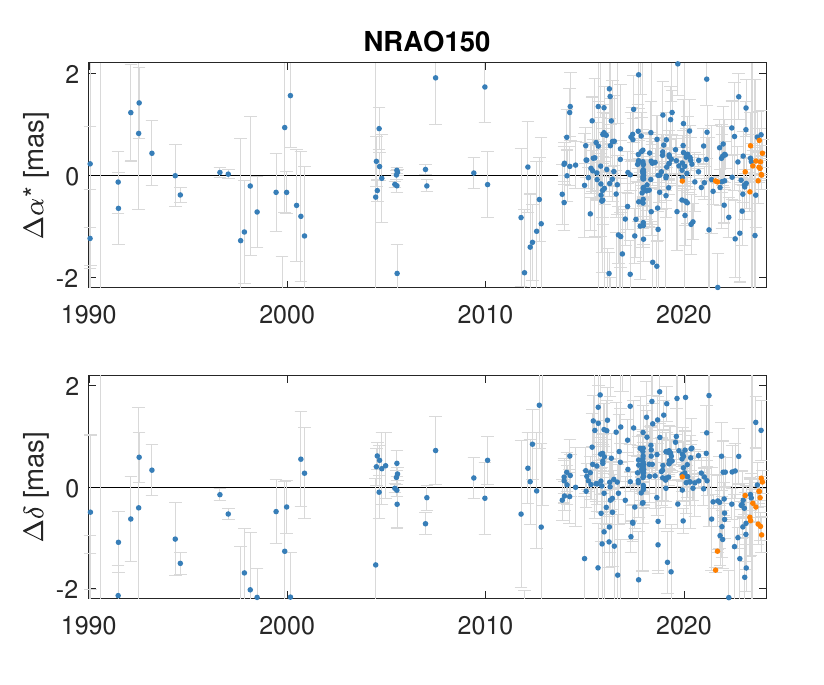}
   \includegraphics[clip,  trim=7cm 5cm 7cm 5cm,width=0.49\hsize]{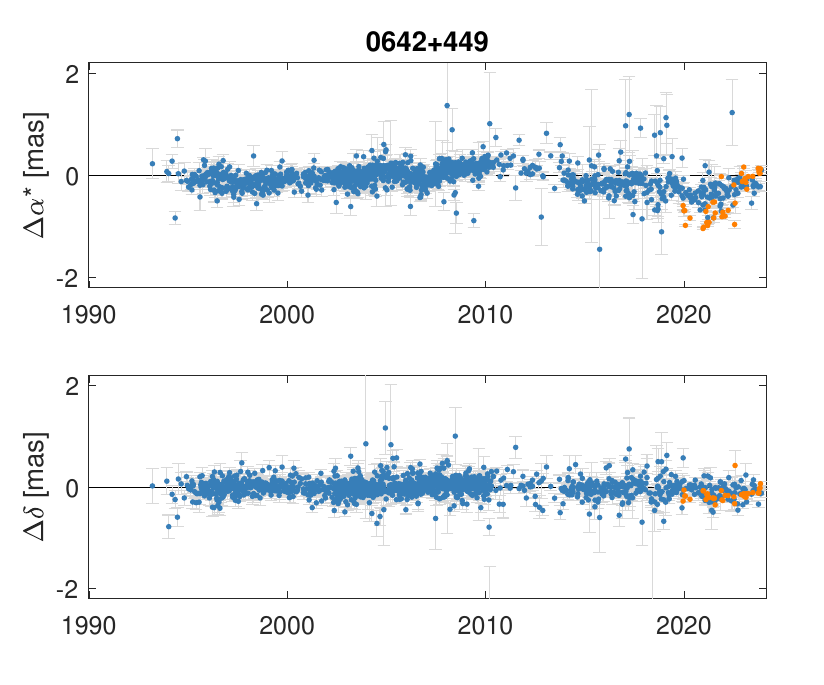}
      \caption{Session-wise position of sources 0119+115, 0229+131, 0355+508 (NRAO150), and 0642+449 with respect to ICRF3-SX from S/X sessions (blue points) and from VGOS sessions (orange points) with more than 20 good observations of the given source in the session. The error bars are formal errors from the single-session solutions.}
         \label{fig:souout}
\end{figure*}

For the four most observed ones,  0119+115, 0229+131, 0355+508 (NRAO150), and 0642+449, we plotted the time series of their position estimates since 1990 from S/X sessions and available VGOS sessions with respect to their ICRF3-SX coordinates (Fig.~\ref{fig:souout}). We plotted only estimates from sessions where the source has more than 20 good observations. In these single-session solutions, we estimated the coordinates of the respective sources without any constraints and we used the same parameterization for the S/X and VGOS sessions. The visual inspection reveals that the measured location of the radio sources is variable over time and this evolution follows a similar pattern from both S/X and VGOS sessions, although with different intensities. To characterize the scatter of the position estimates over the same time span, we computed the weighted standard deviation (wstd) and the weighted Allan deviation \citep[wadev;][]{Malkin2008} of the estimates for both datasets from sessions since 2019 (Table~\ref{tab:souwstd}). In addition, we inspected the available images of the sources.
The X-band image of 0119+115 shows a sub-structure in the north-south direction which projects into the declination estimates of the source. The wstd of $\Delta\delta$ (560~\textmu as for S/X data, and 757~\textmu as for VGOS) is a factor of two higher than for the respective $\Delta\alpha^*$ estimates.
Images of 0229+131 and 0642+449 reveal a sub-structure in the east-west direction which propagates into the $\alpha^*$ estimates. The declination estimates of 0229+131 from VGOS are rather effected by some outlier sessions (Fig.~\ref{fig:souout}). 

In addition, we computed the difference between the mean source coordinates from S/X and VGOS single-session estimates weighted by their formal errors from 2019.0 to 2024.0. By excluding the S/X sessions before 2019.0 we ruled out the variations in position estimates that the radio sources exhibited before the beginning of VGOS measurements, but which---of course---influenced the ICRF3-SX positions. Indeed, we see that the removal of data prior to 2019.0 reduces the estimated position offset between the S/X and VGOS data by several tens of percent for the discussed outlier sources 0119+115, 0229+131, and 0642+449 in $\Delta\alpha^*$ of 55\%, 34\%, 43\%, and in $\Delta\delta$ of 95\%, 65\%, and 58\% for the respective three sources. 

Source NRAO150 is special due to its location in the Galactic plane (Galactic latitude = $-1.6\degr$) where the interstellar medium of our Galaxy is broadening the signal at the discussed radio frequencies which has the consequence that most of its flux gets resolved out on very long baselines. Therefore, there are observations from short baselines only which causes the large scatter with large uncertainties of the NRAO150 position estimates in both coordinates over the entire observation history. Therefore, when we compared the coordinate difference between \crfname{} and ICRF3-SX versus the weighted mean difference from 2019.0 to 2024.0 we got similar values: $-43$~\textmu as versus $-50$~\textmu as for $\Delta\alpha^*$, and $-688$~\textmu as versus $-648$~\textmu as for $\Delta\delta$. This suggests that the main error part for NRAO150---the interstellar medium scattering---does not change over time.

\subsection*{Vector spherical harmonics}

\begin{table}
\tiny
\caption{VSH parameters up to degree and order two for \crfname{} with respect to ICRF3-SX and VIE2023sx.}
\label{tab:souVSH}       
\begin{tabular}{l|S@{\hspace{-0.55cm}}c@{\,}S   S@{\hspace{-0.55cm}}c@{\,}S S@{\hspace{-0.55cm}}c@{\,}S   S@{\hspace{-0.55cm}}c@{\,}S}
\hline\hline\noalign{\smallskip}
   &\multicolumn{3}{c}{ICRF3-SX}  & \multicolumn{3}{c}{VIE2023sx}  &\multicolumn{3}{c}{ICRF3-SX}& \multicolumn{3}{r}{VIE2023sx}\\
$D_n$      &\multicolumn{3}{c}{< 15}  & \multicolumn{3}{c}{< 15} & \multicolumn{3}{c}{< 5} & \multicolumn{3}{c}{< 5}\\
No. sources & \multicolumn{3}{c}{408} & \multicolumn{3}{c}{410} & \multicolumn{3}{c}{333} & \multicolumn{3}{c}{327} \\
\hline
$R_1$ &  +56  &$\pm$&  16  &  +51 &$\pm$& 16 &   +22 &$\pm$& 19   & +19 &$\pm$& 10  \\
$R_2$ &  -26  &$\pm$& 17   &   -0 &$\pm$& 16 &   -28 &$\pm$& 11   &+0 &$\pm$& 10 \\
$R_3$ &  +24 &$\pm$&  9    &  +38 &$\pm$&  9 &   +21 &$\pm$& 6   & +27 &$\pm$& 6   \\
 \noalign{\smallskip}
$D_1$ &   -22 &$\pm$&  16  &  -19 &$\pm$& 15 &   -27 &$\pm$& 11   & -20 &$\pm$& 10    \\
$D_2$ &   -61 &$\pm$&  16  &  -47 &$\pm$& 15 &  -25 &$\pm$& 10    & -28 &$\pm$& 9 \\
$D_3$ &     -1 &$\pm$&  13  & +27 &$\pm$& 12 &  +1 &$\pm$& 8     & +26 &$\pm$& 8  \\
 \noalign{\smallskip}
$a_{2,0}^{e}$    & -28 &$\pm$&  13 &   -52 &$\pm$& 13  &  -17 &$\pm$& 9          & -30 &$\pm$& 8   \\
$a_{2,0}^{m}$   &  +18 &$\pm$&  11 &   +14 &$\pm$& 11 &  +16 &$\pm$& 7           & +1 &$\pm$& 7    \\
$a_{2,1}^{e,Re}$ & -59 &$\pm$&  16&    -54 &$\pm$& 15 & -28 &$\pm$& 11           & -25 &$\pm$& 10   \\
$a_{2,1}^{e,Im}$ & +35 &$\pm$&  17&    +28 &$\pm$& 16 &  +21 &$\pm$& 11          & +22 &$\pm$& 10   \\
$a_{2,1}^{m,Re}$ & -36 &$\pm$&  16&    -25 &$\pm$& 15 &   -10 &$\pm$& 10          & -11 &$\pm$& 10   \\
$a_{2,1}^{m,Im}$ & -21 &$\pm$&  17&    -14 &$\pm$& 16 &  -29 &$\pm$& 11          & -25 &$\pm$& 10   \\
$a_{2,2}^{e,Re}$ &  +11 &$\pm$&  6&   +9 &$\pm$&  6  &  +4 &$\pm$& 4             &  +1 &$\pm$&  4  \\
$a_{2,2}^{e,Im}$ &  -1 &$\pm$&  6&    -0 &$\pm$&  6  & +3 &$\pm$& 4             & +2 &$\pm$&  4   \\
$a_{2,2}^{m,Re}$ &  -1 &$\pm$&  7&   -6 &$\pm$&  7   & -3 &$\pm$& 5            & -1 &$\pm$&  4   \\
$a_{2,2}^{m,Im}$ &  +3 &$\pm$&  7&   +3 &$\pm$&  7   &  +4 &$\pm$& 5            & +3 &$\pm$&  4  \\
\hline
\end{tabular}
\tablefoot{
Units are \textmu as. The second and third columns show comparisons for sources with normalized angular separation $D_n < 15$. In the last two columns the comparison for sources with $D_n < 5$ with respect to ICRF3-SX (85 outliers) and VIE2023sx (91 outliers) is shown.}
\end{table}

After excluding the ten outlier sources from the dataset ($D_n > 15$), we computed for the remaining 408 common sources the vector spherical harmonics \citep[VSH;][]{Mignard2012} up to degree and order two (i.e., 16 parameters) to examine possible systematic differences with respect to ICRF3-SX (Table~\ref{tab:souVSH}, Col.~2). The largest systematic effect reaches 61~\textmu as. There are three parameters with absolute values around 60~\textmu as: $R_1 = +56\pm~16$~\textmu as, 
$D_2 = -61~\pm~16$~\textmu as,
$a_{2,1}^{e,Re} = -59~\pm~16$~\textmu as.
If we compare the \crfname{} to the VIE2023sx which includes VLBI sessions until 2024.0, the VSH are reduced only marginally (the largest absolute value of the VSH is $a_{2,1}^{e,Re} = -54~\pm~15$~\textmu as). We note that there are correlations between the solved 16 VSH. Solving for just three rotations between \crfname{} and ICRF3-SX (including the 408 sources with $D_n < 15$), one gets following values: $R_1 = -8~\pm~10$~\textmu as,  $R_2 = +12~\pm~10$~\textmu as, $R_3 = -16~\pm~8$~\textmu as. 
If we lower the limit for outlier detection to $D_n > 5$, the largest absolute values of the 16 VSH with respect to both, ICRF3-SX (333 radio sources) and VIE2023sx (327 radio sources), are below 30~\textmu as suggesting that the VSH distortion are due at least in part to source structure in the outlier sources.

\subsection{Assessment of the estimation of path delays in the neutral atmosphere in VGOS CRF solutions}
\label{sec_compare_zwd}
One of the currently discussed topics in the geodetic analysis of the VGOS sessions is the optimal interval for estimation of the tropospheric parameters since the troposphere is one of the dominant error sources which influence the VLBI measurements. 
As an example, let us assume a typical wind speed of 10~m~s$^{-1}$ and an elevation of $6\degr$. Then the time for a wet troposphere at a height of 1--2~km to be carried over the site from zenith to the horizon is the time to travel 10--20~km (i.e., troposphere height divided by the sine of elevation) which equals to 16--32 minutes.
The VGOS-OPS have observed typically around 40 scans per hour, whereas the scheduling approach of VGOS-R\&D led to a 2.3-fold increase in the number of scheduled scans per station in 2022 \citep{Schartner2024}. The high cadence of observations during a VGOS session enables estimation of the wet delay in the zenith direction (zwd) and of the horizontal tropospheric gradients in dense intervals. The horizontal gradients describe the deviation from the azimuthal symmetry of the troposphere and we took the mean site gradients computed from Data Assimilation Office \citep[DAO;][]{MacMillan1997} as a priori values. 

\begin{figure}
   \centering
   \includegraphics[clip, trim=4cm 14cm 4cm 8.5cm, width=\hsize]{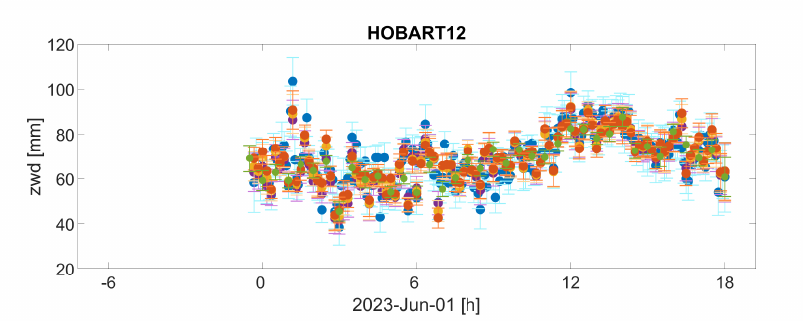}\\
   \includegraphics[clip, trim=4cm 14cm 4cm 8.5cm, width=\hsize]{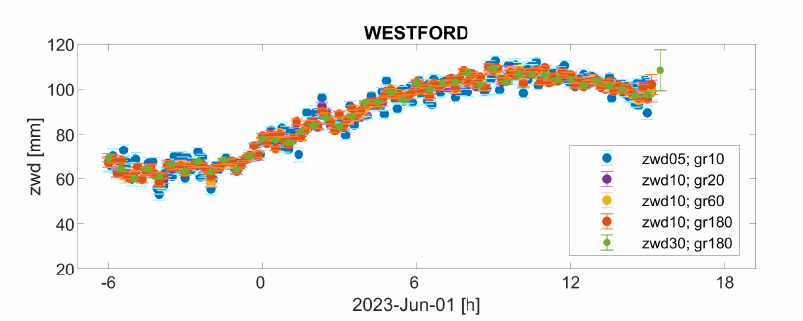}
      \caption{Time series of estimated zenith wet delays at stations Hb (upper plot) and Wf (lower plot) in session VR2302. Five different solutions were computed with varying estimation intervals for zwd and tropospheric gradients (interval lengths are given in the legend in minutes).}
         \label{zwdintervals}
\end{figure}

In our default solution \crfname{} we estimated the zwd each 30 min and tropospheric gradients every 3 hours. The parameterization was done with piece-wise linear offsets with relative constraints of 15~mm and 0.5~mm between the consecutive zwd and gradient estimates, respectively.
To examine the impact of shorter estimation intervals of zwd and gradients on the CRF, we ran the following four test solutions: first, we reduced the zwd estimation interval to 10 min and set the gradient estimation interval in the following three solutions zwd10gr180, zwd10gr60, and zwd10gr20 to 180~min, 60~min, and 20~min, respectively. Finally, we computed a solution zwd05gr10 with 5~min zwd interval and 10~min gradient intervals. As an example, we show zwd estimates (Fig.~\ref{zwdintervals}) at antennas Hb and Wf during the VGOS-R\&D session VR2302.

\begin{figure}
   \centering
   \includegraphics[clip, trim=4cm 8.5cm 4cm 9cm, width=\hsize]{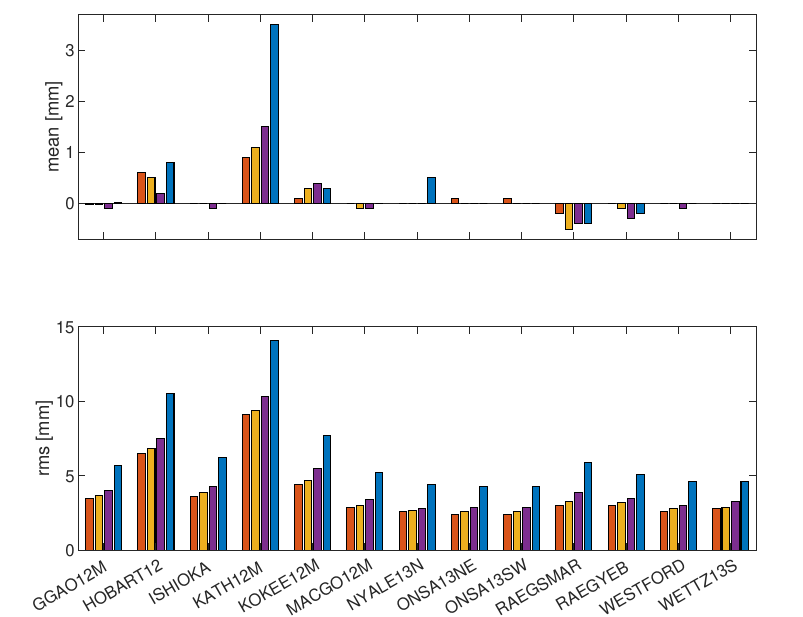}\\
      \caption{Mean and rms for the differences in the zwd time series with respect to default solution zwd30grd180 at individual stations from all VGOS sessions. The legend for the four solutions is given in Fig.~\ref{zwdintervals}.}
         \label{zwdstat}
\end{figure}

We characterize the difference between the zwd estimates from the four solutions with respect to \crfname{} (i.e., zwd30gr180) with the mean difference and the rms (Fig.~\ref{zwdstat}). The statistics were computed from all VGOS sessions and the zwd values were interpolated to the five-minute intervals that correspond to the solution zwd05gr10. The mean difference of the zwd estimates at all telescopes is less than 0.5~mm, except for Hb and Ke. For Ke the largest mean difference in the zwd is 3~mm between solutions zwd05gr10 and zwd30gr180. For Ke there were a limited number of sessions for the statistics since it participated in only five VGOS sessions (Fig.~\ref{Ant_VGOSCRF}). Ke and Hb are the only stations in the southern hemisphere, meaning that the mutual observations with telescopes in the northern hemisphere do not allow uniform sky coverage, which limits their capability to estimate the tropospheric parameters. 

The rms values of zwd differences (Fig.~\ref{zwdstat}, lower panel) from solutions with zwd interval 10 min with respect to zwd30gr180 are between 3--4~mm at the majority of stations. The variation of gradient intervals does not affect the rms for zwd significantly. The rms of zwd differences for a solution with short zwd intervals of 5~min with respect to the default solution is about 1~mm larger than from the other solutions. The exceptions are again telescopes Hb and Ke, where the rms with respect to default are twice to three times larger than at other stations. We also examined the horizontal gradients from these four solutions comparing them to zwd30gr180 where the statistics show that differences in the mean and rms at all stations are below 1~mm.

\begin{figure}
   \centering
   \includegraphics[clip, trim=3cm 8cm 4cm 9cm, width=\hsize]{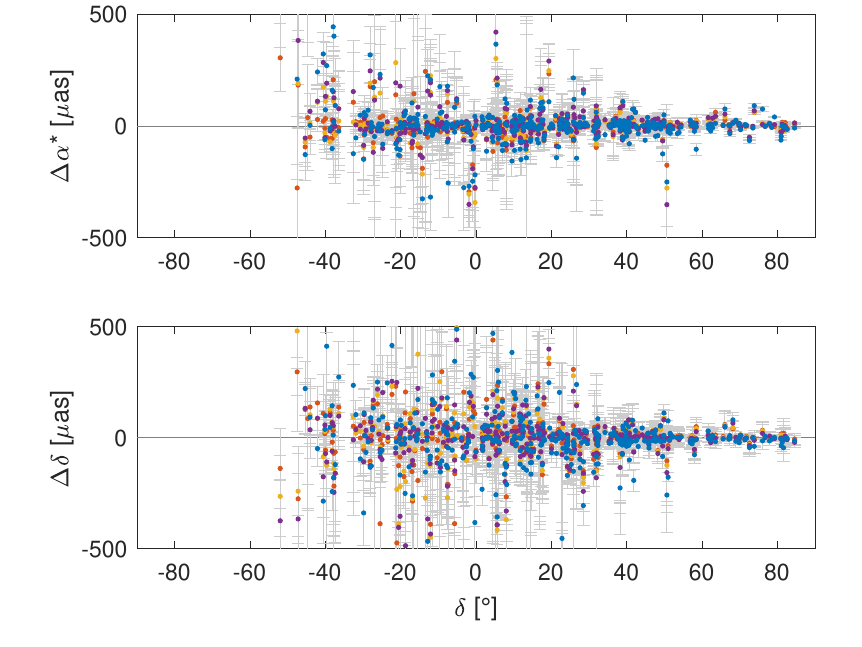}\\
      \caption{Differences in $\alpha^*$ and $\delta$ for four test CRFs with varied estimation intervals for zwd and tropospheric gradients with respect to default \crfname{} (zwd30gr180). Only sources with more than 25~observations are plotted. The legend for the four solutions is given in Fig.~\ref{crf_tropo_diff_sigma}.}
         \label{crf_tropo_diff}
\end{figure}

\begin{figure}
   \centering
   \includegraphics[clip, trim=3cm 8cm 4cm 9cm, width=\hsize]{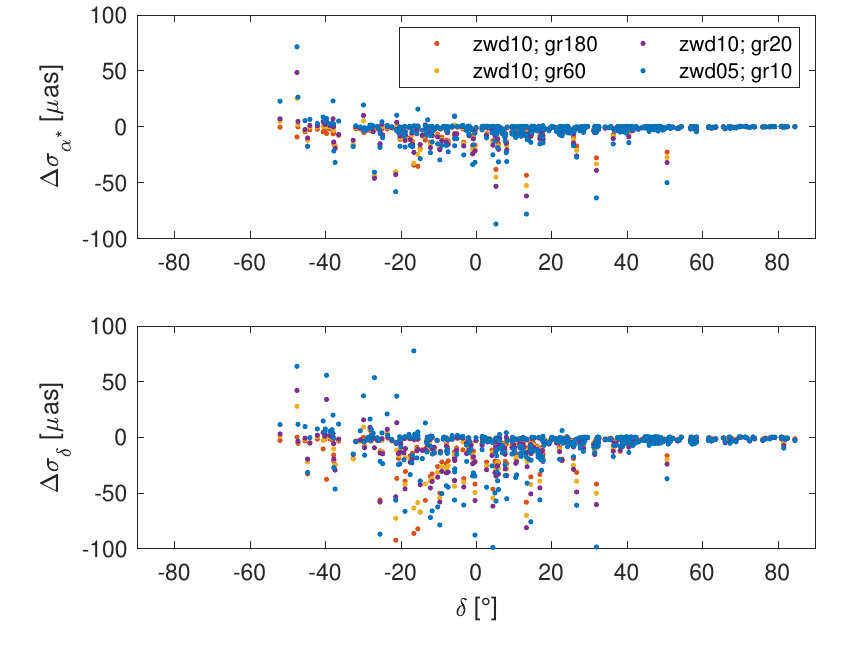}\\
      \caption{Differences in formal errors of $\alpha^*$ and $\delta$ for four test CRFs with varied estimation intervals for zwd and tropospheric gradients with respect to default \crfname{} (zwd30gr180). Only sources with more than 25~observations are plotted.}
         \label{crf_tropo_diff_sigma}
\end{figure}

Focusing on the estimated CRFs in these solutions, the differences in source positions as a function of declination from the four test solutions with respect to the default zwd30gr180 are plotted in Fig.\ref{crf_tropo_diff} and the corresponding formal errors from the four solutions with respect to the formal errors of the default CRF in Fig.~\ref{crf_tropo_diff_sigma}. In both figures only sources with more than 25~observations are plotted (cf. Fig.~\ref{sched_analyse_scanobs}). In Table~\ref{tab:zwd_diffsigmaCRF} we summarize the mean of the formal error differences with respect to solution zwd30gr180 individually for three groups of sources dependent on their declination, with the empirical declination limits $-30\degr$ and $+20\degr$. The larger position scatter between the CRFs is observed for sources in the southern hemisphere up to $\delta$ about $+20\degr$ (Fig.\ref{crf_tropo_diff}, lower panel). The negative values of differences in formal errors (Fig.~\ref{crf_tropo_diff_sigma}) in the declination range approximately between $-30\degr$ and $+20\degr$ show that for most of the sources especially the declination estimates have lower formal errors in the test solutions with shorter estimation intervals of the zwd than in the default solution (e.g., the mean difference between declination formal errors for zwd05gr10 with respect to zwd30gr180 is $-14$~\textmu as; see Table~\ref{tab:zwd_diffsigmaCRF}). This can be explained by the fact that these sources were observed at lower elevation angles by the north antennas. The more frequent zwd estimates at the antennas with good sky coverage allow one to estimate the declination component of the ``low elevation sources'' with a formal error lower by several tens of microarcseconds for several sources. For the northern sources ($\delta > +20\degr$) there is no such improvement in the declination formal errors (mean difference is $-4$~\textmu as between zwd05gr10 and zwd30gr180), suggesting that the more frequent zwd estimates have no significant effect on the estimated position of sources observed at high elevation angles. On the other hand, for the majority of southern sources with declination lower than $-30\degr$ their formal errors of the position estimates are higher in the solution zwd05gr10 ($5$~\textmu as difference for both coordinates). The position of these sources is sensitive to the zwd estimates at stations Hb and Ke which lack sufficient sky coverage for frequent zwd and horizontal gradient estimates.

\begin{table}
\caption{Mean difference between formal errors in right ascension and declination estimates for the four CRF solutions with respect to zwd30gr180. }
\label{tab:zwd_diffsigmaCRF}       
\tabcolsep=0.17cm
\begin{tabular}{lrrrr}
\hline
\hline\noalign{\smallskip}
  &zwd10gr180&zwd10gr60&zwd10gr20&zwd05gr10\\
\noalign{\smallskip}\hline
\multicolumn{5}{c}{$\delta < -30\degr$ }\\
 $\Delta\sigma_{\alpha^*}$ & $-6$ &  $-2$  &  $1$  &     $5$ \\
$\Delta\sigma_{\delta}$ & $-9$    &     $-4$   &      $-1$     &    $ 5$\\
\multicolumn{5}{c}{$-30\degr <\delta < 20\degr$ }\\
 $\Delta\sigma_{\alpha^*}$ &  $-5$   &      $-5$  &      $ -5$     &    $-7$\\
$\Delta\sigma_{\delta}$ & $-13$   &    $ -12$  &      $-12$    &   $ -14$\\
\multicolumn{5}{c}{$\delta > 20\degr$ }\\
$\Delta\sigma_{\alpha^*}$ & $-2$    &      $-2$  &        $-2$    &      $-3$\\
$\Delta\sigma_{\delta}$ & $-3$    &     $-3$     &    $-4$   &     $ -4$\\
\hline\noalign{\smallskip}
\end{tabular}
\tablefoot{The mean value is computed for three groups of sources depending on their declination. Only sources with more than 25 observations are included in the statistics. The units are \textmu as.}
\end{table}

In Table~\ref{tab:crftropo} we summarize the statistics on the source coordinate estimates (mean and rms) for the four solutions with respect to zwd30gr180. Here we computed the mean and rms for three subgroups of sources depending on the minimal number of observations in the CRF (50, 25, and 2). In contrast to the pure mean and rms which differ in dependence on the sources in the group, the weighted mean and weighted rms are independent of the selection of sources.
There is no significant bias (not higher than 10~\textmu as) between the four solutions and zwd30gr180, if computed for sources with a minimum of 25 observations. If also sources with 2--25 observations are included in the statistics, the mean difference in declination between zwd05gr10 and zwd30gr180 is 38~\textmu as. It is important to consider that sources with a limited number of observations are in general sensitive to the parameterization options (such as the interval for the zwd and gradient estimation) during a CRF estimation. This is of importance especially when fixing the coordinates of the radio sources to a given CRF in the VLBI analysis.

\begin{table*}
\caption{Statistics for difference in $\Delta\alpha^*$ and $\Delta\delta$ for four CRF solutions with respect to \crfname{} (i.e., zwd30gr180).}
\label{tab:crftropo}       
\begin{tabular}{lrr rr rr rr  rr rr  rr rr}
\hline
\hline\noalign{\smallskip}
 &
  \multicolumn{4}{c}{} &
 \multicolumn{4}{c}{nobs>50} &
  \multicolumn{4}{c}{nobs>25}  &
 \multicolumn{4}{c}{nobs>2}\\
 & \multicolumn{2}{c}{w.mean} &
 \multicolumn{2}{c}{wrms} &
 \multicolumn{2}{c}{mean} &
 \multicolumn{2}{c}{rms} &
  \multicolumn{2}{c}{mean} &
 \multicolumn{2}{c}{rms} &
  \multicolumn{2}{c}{mean} &
 \multicolumn{2}{c}{rms} \\
 solution& ${\alpha^*}$& $\delta$&$\alpha^*$& $\delta$
 &$\alpha^*$& $\delta$&$\alpha^*$& $\delta$ 
 &$\alpha^*$& $\delta$&$\alpha^*$& $\delta$ 
 &$\alpha^*$& $\delta$&$\alpha^*$& $\delta$ \\
\noalign{\smallskip}\hline
zwd10gr180&0 & 1    & 11 & 20  
& 2& 11& 44&101 
&2&0   &59&170 
&7& $-22$ &185 &517
\\
zwd10gr60 & 0& 0    & 12 & 21  
& 6 &9  & 54 & 121
&6&4   &77&163  
&20&6&252&823
\\
zwd10gr20 & $-1$ & $-1$ & 13 & 23
& 9 & 7 & 64 & 144
&9&7 &100&192 
&16&$-13$&274&899
\\
zwd05gr10 &$-5$  &$-6$ & 25  & 34  
& 7 & 6 & 112 & 185
&10&9 &132&274 
&34&38&569&1752
\\
\hline\noalign{\smallskip}
\end{tabular}
\tablefoot{The mean and rms were calculated for three subgroups of sources corresponding to the minimum number of observations (50, 25, and 2). All values are in \textmu as.}
\end{table*}

\subsection{Evaluation of \crfname{} in geodetic analysis}
\label{sec_eval_analysis}
The standard procedure for the estimation of all five EOP and antenna positions in the operational geodetic VLBI analysis is to fix the source positions to their a priori coordinates given in the CRF catalog. Up to now, a priori catalogs estimated from S/X measurements, such as ICRF3-SX, have been taken as standard in the analysis of VGOS sessions within the IVS. Recently, \citet{Diamantidis2023} investigated whether the EOP repeatabilities are related to the CRF definition. They constructed a VGOS-adjusted ICRF3-SX catalog from a weighted mean of source estimates which they obtained by favoring source coordinate over EOP corrections in a Kalman filter on a session-wise basis, and evaluated it in terms of EOP and station positions. Their results show an improvement in station position and baseline length repeatabilities in both on the level of 0.3--0.5~mm for all but the
east component, and they obtain a lower wrms of the EOP estimates by approximately 50--60\%.

We analyzed the full set of VGOS-OPS and VGOS-R\&D sessions until 2024.0 with the standard parameterization (Table~\ref{tab:solparam}) and created two groups of session-wise solutions: S1 with source coordinates fixed to ICRF3-SX, and S2 with source coordinates fixed to \crfname{}. The a priori station positions were ITRF2020 (Nn is the exception) and the NNT/NNR condition was imposed session-wise with respect to all ITRF2020 stations to follow the current standard way of VGOS analysis (Table~\ref{tab:s1s4}). Additionally, two other solutions (S3 and S4) with updated a priori station positions from VIE2023 TRF and applied NNT/NNR with respect to VIE2023 TRF on all stations were created.

\begin{table}
\tiny
\caption{Differences between single-session solutions S1--S4.}
\label{tab:s1s4}       
\begin{tabular}{lrr}
\hline
\hline\noalign{\smallskip}
 solution& CRF& TRF \\
\noalign{\smallskip}\hline
S1 & ICRF3-SX fixed & ITRF2020 NNT/NNR\\
S2 & \crfname{} fixed & ITRF2020 NNT/NNR\\
S3 & ICRF3-SX fixed & VIE2023 TRF NNT/NNR\\
S4 & \crfname{} fixed & VIE2023 TRF NNT/NNR\\
\hline\noalign{\smallskip}
\end{tabular}
\end{table}

\subsubsection{Comparison of ITRF2020 and VIE2023 TRF for VGOS telescopes}

VIE2023 TRF is a terrestrial reference frame computed from S/X and VGOS sessions until 2024.0 which comes from an updated global solution described in \citet{Krasna2023}. Therefore, it includes three additional years of VLBI data compared to ITRF2020, which is crucial for determining the positions and velocities of new VGOS antennas.

\begin{figure}
   \centering
   \includegraphics[clip,  trim=4cm 8cm 4cm 9cm, width=\hsize]{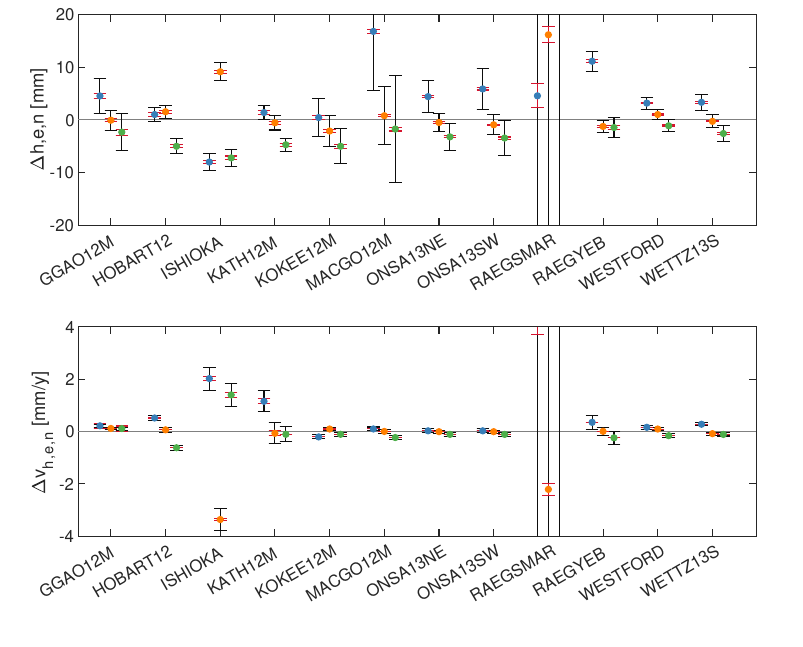}
      \caption{Differences between VIE2023 TRF with respect to ITRF2020 for the  VGOS telescopes. The position offset for the reference epoch 2015.0 (upper panel) and the difference in the linear velocity (lower panel) are shown for the height component as blue dots, for the east component as orange dots, and for the north component as green dots. The VIE2023 TRF uncertainties are plotted as red error bars, and the  ITRF2020 uncertainties as black error bars. The differences at Sa exceed the y-axis limits (see Table~\ref{tab:itrf2020diff} for the exact values).}
         \label{itrf2020HENstat}
\end{figure}

In Fig.~\ref{itrf2020HENstat} we compare VIE2023 TRF with respect to ITRF2020 for the VGOS antennas in terms of the difference in the position offset for the common reference epoch 2015.0 (upper panel) and the difference in the linear velocity (lower panel) in the local coordinate systems: height (h), east (e), north (n). We also depict the reported TRF uncertainties as error bars separately for VIE2023 TRF (in red) and ITRF2020 (in black). The importance of the updated TRF especially for the new VGOS telescopes like Sa (three sessions in ITRF2020) and Mg (seven sessions in ITRF2020) is already discussed in \citet{Krasna2023}. The values for the Sa telescope exceed the y-axis limits in Fig.~\ref{itrf2020HENstat} but they are summarized with all other VGOS telescopes in Table~\ref{tab:itrf2020diff}.

The weighted mean differences in the position offset at epoch 2015.0 computed over the plotted VGOS telescopes are 2.1~mm, 0.6~mm, $-3.5$~mm for the h, e, n components, respectively, and the mean differences for the linear velocity are 0.15~mm/y, 0.02~mm/y, $-0.12$ mm/y. The systematic positive difference between the estimated heights in VIE2023 TRF with respect to ITRF2020 confirms the known discrepancy between the VLBI TRF and ITRF2020 scale \citep{Altamimi23, Krasna2023}.
The systematic negative trend observed for the northern component of the station position differences is likely to be related to differences in the assumed declination of the radio sources (fixing source positions to ICRF3-SX in the IVS contribution to ITRF2020 \citep{Hellmers2022} versus estimated source positions in the VIE2023 global solution). However, this discrepancy needs further investigation.

\begin{table*}
\tiny
\caption{Comparison of VIE2023 TRF and ITRF2020 for the  VGOS telescopes.}
\label{tab:itrf2020diff}       
\begin{tabular}{l|rrrrrrrrr|rrrrrrrrr}
\hline
\hline\noalign{\smallskip}
 &
  \multicolumn{9}{c}{$\Delta$ position  (epoch 2015.0), $\sigma$ VIE2023 TRF,  $\sigma$ ITRF2020}& \multicolumn{9}{c}{$\Delta$ linear velocity, $\sigma$ VIE2023 TRF,  $\sigma$ ITRF2020} \\
  &
\multicolumn{3}{c}{height [mm]}&
\multicolumn{3}{c}{east [mm]}&
  \multicolumn{3}{c}{north [mm]}&
 \multicolumn{3}{c}{height [mm/y]}&
\multicolumn{3}{c}{east [mm/y]}&
  \multicolumn{3}{c}{north [mm/y]} 
 \\
\noalign{\smallskip}\hline
Gs &     4.5 &      0.5 &     3.4    &    -0.1 &      0.3 &     1.9    &    -2.3 &      0.6 &     3.4    &    0.22 &      0.08 &    0.05      &    0.12 &     0.04 &    0.05      &    0.11 &     0.08 &    0.06 \\ 
Hb &     1.0 &      0.3 &     1.4    &     1.5 &      0.3 &     1.3    &    -5.0 &      0.3 &     1.4    &    0.52 &      0.03 &    0.09      &    0.06 &     0.03 &    0.08      &   -0.62 &     0.03 &    0.10 \\ 
Is &    -8.0 &      0.3 &     1.7    &     9.1 &      0.3 &     1.7    &    -7.2 &      0.4 &     1.7    &    2.02 &      0.05 &    0.44      &   -3.36 &     0.05 &    0.43      &    1.40 &     0.05 &    0.44 \\ 
Ke &     1.4 &      0.3 &     1.4    &    -0.6 &      0.3 &     1.4    &    -4.7 &      0.3 &     1.2    &    1.15 &      0.05 &    0.40      &   -0.07 &     0.05 &    0.40      &   -0.11 &     0.04 &    0.28 \\ 
K2 &     0.4 &      0.3 &     3.5    &    -2.1 &      0.3 &     2.9    &    -5.0 &      0.4 &     3.3    &   -0.21 &      0.02 &    0.08      &    0.09 &     0.02 &    0.07      &   -0.11 &     0.02 &    0.07 \\ 
Mg &    16.8 &      0.3 &    11.2    &     0.8 &      0.2 &     5.5    &    -1.8 &      0.4 &    10.1    &    0.10 &      0.02 &    0.08      &    0.00 &     0.01 &    0.06      &   -0.23 &     0.01 &    0.07 \\ 
Oe &     4.4 &      0.2 &     3.1    &    -0.5 &      0.1 &     1.6    &    -3.2 &      0.2 &     2.6    &    0.03 &      0.01 &    0.07      &   -0.01 &     0.01 &    0.06      &   -0.11 &     0.01 &    0.06 \\ 
Ow &     5.8 &      0.2 &     4.0    &    -0.9 &      0.1 &     1.9    &    -3.5 &      0.2 &     3.3    &    0.03 &      0.01 &    0.07      &   -0.01 &     0.01 &    0.06      &   -0.11 &     0.01 &    0.06 \\ 
Sa &     4.5 &      2.3 &   390.8    &    16.1 &      1.5 &   384.7    &   -53.4 &      2.3 &   388.0    &    4.01 &      0.30 &   96.92      &   -2.21 &     0.20 &   95.63      &    9.42 &     0.30 &   96.49 \\ 
Yj &    11.1 &      0.3 &     1.9    &    -1.3 &      0.2 &     1.2    &    -1.5 &      0.3 &     1.9    &    0.35 &      0.04 &    0.26      &    0.01 &     0.02 &    0.14      &   -0.24 &     0.04 &    0.26 \\ 
Wf &     3.2 &      0.1 &     1.1    &     1.0 &      0.1 &     1.0    &    -1.1 &      0.2 &     1.1    &    0.16 &      0.01 &    0.07      &    0.09 &     0.01 &    0.06      &   -0.16 &     0.01 &    0.07 \\ 
Ws &     3.3 &      0.2 &     1.5    &    -0.3 &      0.1 &     1.2    &    -2.6 &      0.2 &     1.5    &    0.28 &      0.01 &    0.06      &   -0.08 &     0.01 &    0.05      &   -0.11 &     0.01 &    0.06 \\ 
\hline\noalign{\smallskip}
\end{tabular}
\tablefoot{Differences ($\Delta$, VIE2023 TRF minus ITRF2020) in position at the reference epoch 2015.0 and in linear velocity are expressed in the local height, east, north system. Each coordinate component is followed with the relevant uncertainty reported separately for VIE2023 TRF and for ITRF2020.}
\end{table*}

\subsubsection{Baseline length and station position}

\begin{figure}
   \centering
   \includegraphics[clip, trim=4cm 9.5cm 4cm 9cm,  width=\hsize]{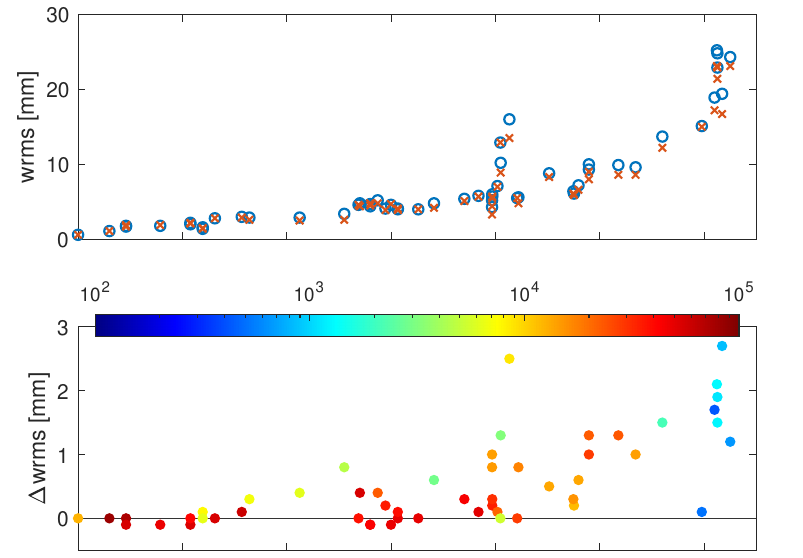}
   \includegraphics[clip, trim=4cm 14cm 4cm 8cm,  width=\hsize]{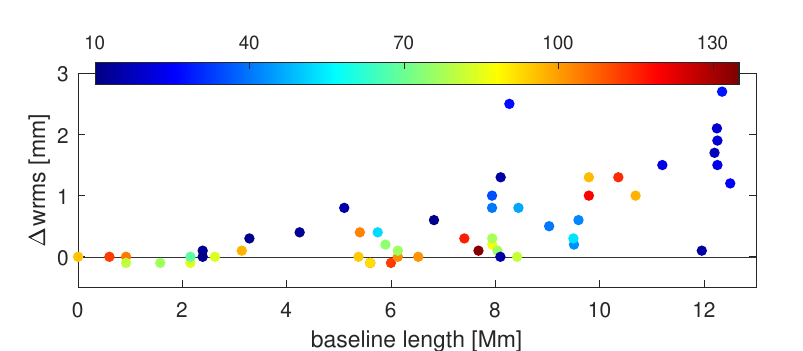}   
      \caption{Baseline length repeatability (upper panel) from all VGOS sessions in solution S1 (blue circles) and in solution S2 (orange x-signs). The lower plots show the difference in wrms computed as S1 minus S2 with the total number of observations (middle panel) and with the number of baseline length determinations (bottom panel) color-coded for each baseline. Only baselines included in more than ten sessions are shown.}
         \label{wrmsbas}
\end{figure}

\begin{figure}
   \centering
   \includegraphics[clip, trim=4cm 14cm 4cm 9cm,  width=\hsize]{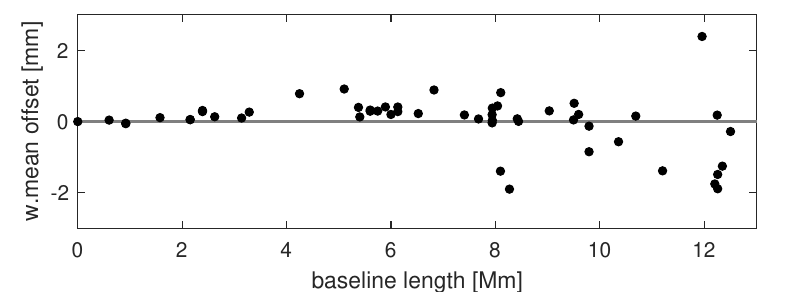}   
      \caption{Weighted mean of differences (S1 minus S2) between the absolute baseline length time series obtained from VGOS sessions. Only baselines observed in more than ten sessions are shown.}
         \label{doffsetbas}
\end{figure}

Figure~\ref{wrmsbas} depicts the baseline length repeatability (computed as wrms of estimated baseline length from all VGOS sessions) from solutions S1 and S2 (upper panel), and the difference in wrms computed as S1 minus S2 in the two lower panels. We show only baselines that were observed in more than ten sessions. As the length of the baselines gets larger, the improvement in the wrms for individual baselines in solution S2 over S1 increases. This change is described with a fitted linear regression line as $\Delta \mathrm{wrms} = 0.14 \cdot L - 0.42$  [mm] with $L$ being the length of a baseline in megameters. The largest improvement of 2.7~mm is seen for the baseline Hb-Wf (\num{12350}~km). The mean improvement computed over extremely long baselines only (i.e., longer than \num{12000}~km) is 1.9~mm. In this VGOS dataset, all baselines longer than 11 thousand kilometers scheduled in more than five sessions include Hb and due to the geometry, all those common observations are conducted at very low elevation angles. In general, the number of observations for those extremely long baselines is lower than for the others (see color-codes in the middle plot of Fig.~\ref{wrmsbas}). The number of baseline length determinations in our current dataset is depicted in the lower plot of Fig.~\ref{wrmsbas}. Other possible causes for the increasing difference in baseline length wrms with increasing baseline length, such as the larger projection of the frequency-dependent offset between the ICRF3-SX and \crfname{} positions on the long baselines, require further investigation. In addition to the baseline length scatter represented by the wrms, we computed the weighted mean between the absolute baseline length time series differences from solution S1 with respect to S2 (Fig.~\ref{doffsetbas}). For most of the baselines the absolute value of the weighted mean of the baseline length differences is lower than 1~mm. Nevertheless, the weighted mean offsets for long baselines especially with Hb reach up to 2.4~mm.

\begin{figure}
   \centering
   \includegraphics[clip,  trim=4cm 8cm 4cm 9cm, width=\hsize]{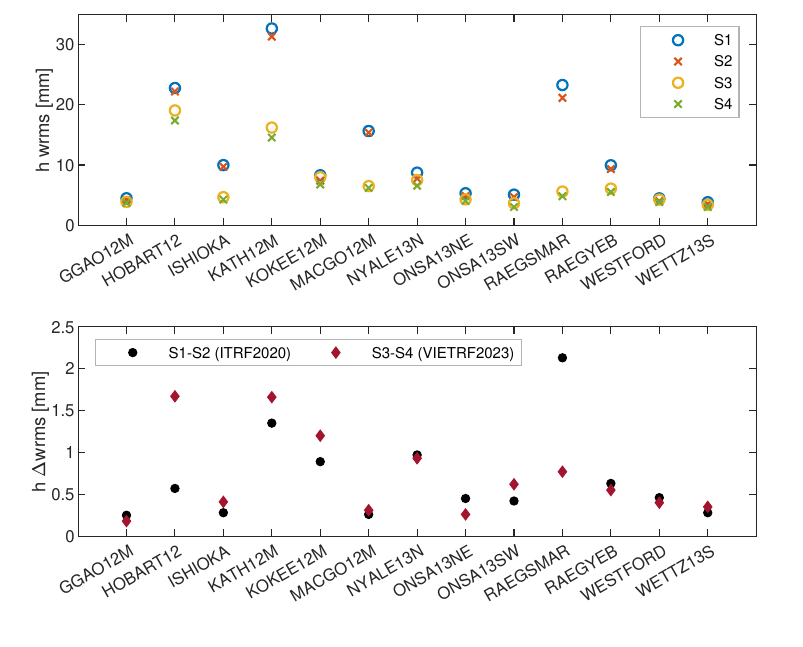}
      \caption{Wrms of the height position time series from all VGOS sessions in solution S1 (ICRF3, ITRF2020) and in solution S2 (\crfname{}, ITRF2020). In addition, wrms from solutions with VIE2023 TRF a priori are shown: S3 (ICRF3, VIE2023 TRF) and S4 (\crfname{}, VIE2023 TRF). The lower plot depicts the difference in wrms computed as solutions with fixed ICRF3 minus solutions with fixed \crfname{}.}
         \label{wrmsHstat}
\end{figure}

\begin{figure}
   \centering
   \includegraphics[clip,  trim=4cm 8cm 4cm 9cm, width=\hsize]{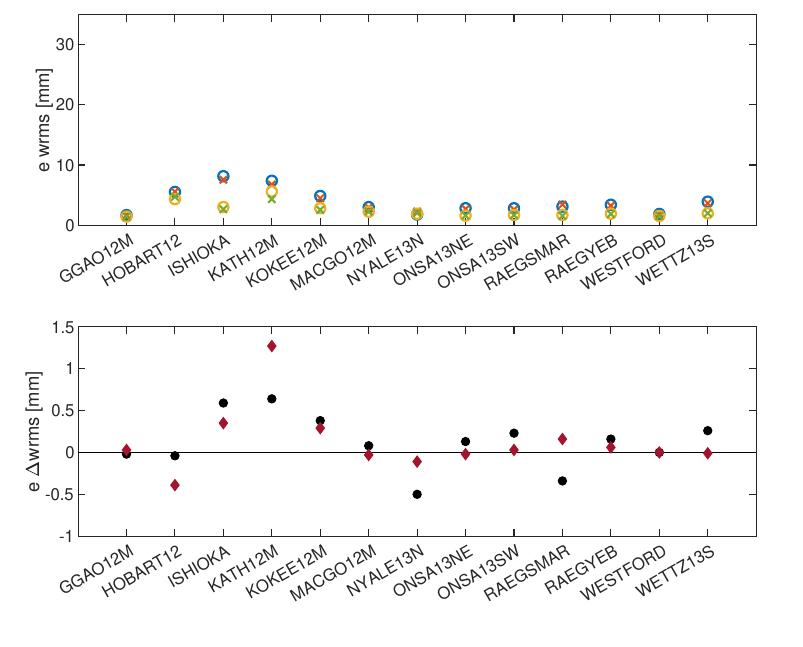}
      \caption{Wrms of the the east component of station position time series from all VGOS sessions (see Fig.~\ref{wrmsHstat} for description).}
         \label{wrmsEstat}
\end{figure}

\begin{figure}
   \centering
   \includegraphics[clip,  trim=4cm 8cm 4cm 9cm, width=\hsize]{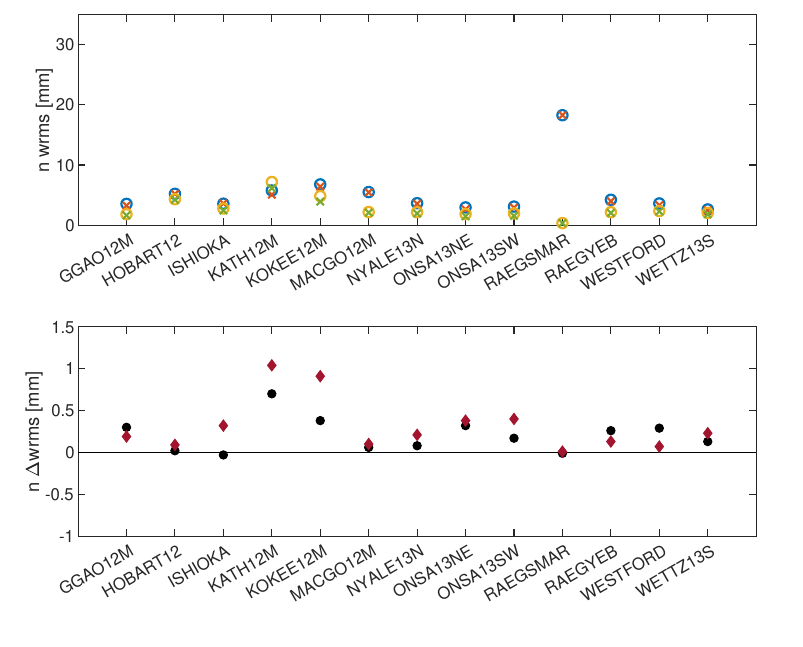}
      \caption{Wrms of the north component of station position time series from all VGOS sessions (see Fig.~\ref{wrmsHstat} for description).}
         \label{wrmsNstat}
\end{figure}

In addition to the baseline length, we present the impact of the two fixed CRFs on the station position time series. The wrms for each of the three components of the local coordinate system h, e, n are shown for each telescope in Figs.~\ref{wrmsHstat}--\ref{wrmsNstat}. In addition to the comparison of the already discussed solutions S1 (blue circles) and S2 (orange x-signs), we computed the wrms from solutions S3 (yellow circles) and S4 (green x-signs), where the only difference is in the a priori TRF. Instead of ITRF2020, we took VIE2023 TRF as a priori TRF in solutions S3 and S4. With VIE2023 TRF a priori, the wrms in height is below 20~mm for all antennas, while that in the horizontal plane is below 10~mm for all antennas. 
The differences in wrms between the solutions S1 and S2 (black dots) and between S3 and S4 (red diamonds) are depicted in the lower plots of Figs.~\ref{wrmsHstat}--\ref{wrmsNstat}. For the height component, there is an improvement in wrms at all stations from fixing to the \crfname{} instead of fixing to the ICRF3-SX. The mean value of $\Delta$wrms is 0.7~mm for the height regardless of the a priori terrestrial frame. The $\Delta$wrms for the east component are in the range from $-0.5$ to 1.5~mm, with the mean value 0.1~mm. For the north component, there is, similar to the height, an improvement in the wrms at all stations by fixing to the \crfname{}. The mean difference is 0.2~mm for ITRF2020 a priori, and 0.3~mm for VIE2023 TRF a priori. We assume that the slightly higher sensitivity to fixing the CRF on the north component than on the east component of station positions is due to the mapping of uncertainties in declination estimates from individual sources into the north component (i.e., latitude) of the stations.

\begin{figure}
   \centering
   \includegraphics[clip,  trim=4cm 9cm 4cm 8.7cm, width=\hsize]{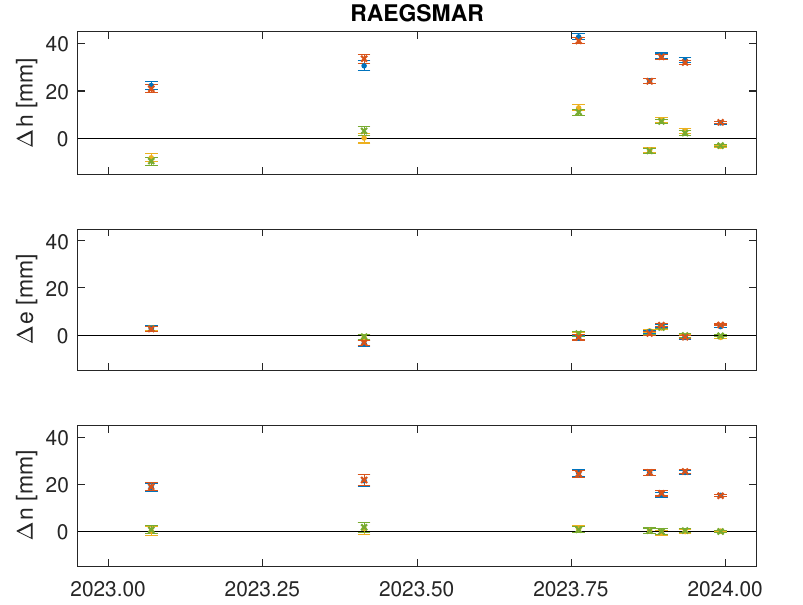}
      \caption{Position estimate time series for station Sa from solutions S1--S4 with respect to a priori TRFs (see Fig.~\ref{wrmsHstat} for legend).}
         \label{dhenSa}
\end{figure}

As an example of station position estimate time series from which the wrms in Figs.~\ref{wrmsHstat}--\ref{wrmsNstat} were computed, we select station Sa (Fig.~\ref{dhenSa}). It participated in seven VGOS sessions till 2024.0 (cf. Fig.~\ref{Ant_VGOSCRF}) and it is the station with largest difference between the reported linear velocity in ITRF2020 and VIE2023 TRF among all VGOS telescopes. The session-wise position estimates in Fig.~\ref{dhenSa} are plotted with respect to the a priori TRF used in the respective solutions, that is with respect to ITRF2020 in S1--S2 and with respect to VIE2023 TRF in S3--S4. The differences between the a priori catalog position of Sa telescope in mid 2023 based on VIE2023 TRF with respect to ITRF2020 are 43~mm, $-5$~mm, 36~mm in the local h, e, n coordinate components. The respective mean differences between the S1 and S3 position estimates are 26.7~mm, 0.0~mm, 20.5~mm (the same values for the mean differences are obtained between the S2 and S4 estimates). The mean estimates with respect to the a priori catalogs are lower than the difference of the a priori positions themselves which is caused by the inclusion of the Sa station in the NNT/NNR condition.


\subsubsection{Earth orientation parameters}

From the solutions S1 and S2, we evaluated the impact of the fixed source positions on EOP. The Earth rotation parameters (polar motion ($\mathrm{x_p, y_p}$) and UT1-UTC) are estimated as piece-wise linear offsets for 0~UTC before the start, inside and after the end of each 24~h session with a relative constraint of 1~mas between the estimates. The nutation is estimated as celestial pole offsets (dX, dY) in the middle of the sessions. 

\begin{figure}
   \centering
   \includegraphics[clip, trim=4cm 14cm 3.5cm 9cm, width=\hsize]{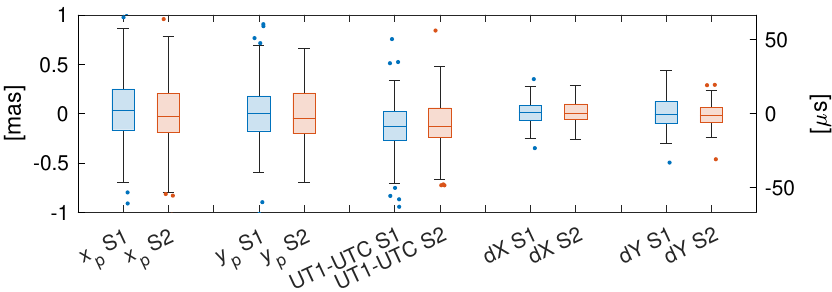}\\
      \caption{Visualization of all five EOP time series from VGOS sessions with respect to IERS 20C04 for solution S1 with fixed ICRF3-SX (in blue) and solution S2 with fixed \crfname{} (in orange).
      The box charts display the median with the lower and upper quartiles. The bar length depicts the minimum and maximum values of the time series that are not outliers. The outliers are shown as dots.}
         \label{boxEOP}
\end{figure}

Figure~\ref{boxEOP} summarizes statistics for all five EOP computed jointly for the VGOS-OPS and VGOS-R\&D sessions from S1 (blue) and S2 (orange) solutions. The statistics are given in terms of a median value with respect to the IERS EOP combined series 20C04.\footnote{\url{https://hpiers.obspm.fr/iers/eop/eopc04/eopc04.1962-now}} The height of the box represents the lower and upper quartiles and the bar length depicts the minimum and maximum values not being outliers. The outliers are here defined as values more than 1.5 times of the interquartile range away from the top or bottom of the box.
In terms of wrms with respect to IERS 20C04, lower values are obtained in solution S2 with fixed \crfname{} for all five EOP. Namely, the wrms computed for detrended time series of $\mathrm{x_p}$, $\mathrm{y_p}$,  UT1-UTC, dX, and dY from solutions S1~/~S2 are  $329~/~293$~\textmu as,   $263~/~237$~\textmu as,   $14~/~13$~\textmu s,    $105~/~97$~\textmu as,  and $123~/~105$~\textmu as, respectively (Table~\ref{tab:EOPstatVO}). This makes an improvement of the wrms in S2 compared to S1 of about 10\% for the first four EOP and of about 15\% for dY.

\begin{figure}
   \centering
   \includegraphics[clip, trim=4cm 14cm 4cm 9cm, width=\hsize]{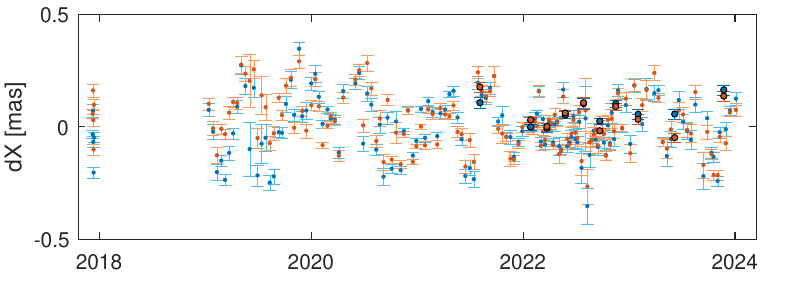}\\
   \includegraphics[clip, trim=4cm 14cm 4cm 9cm, width=\hsize]{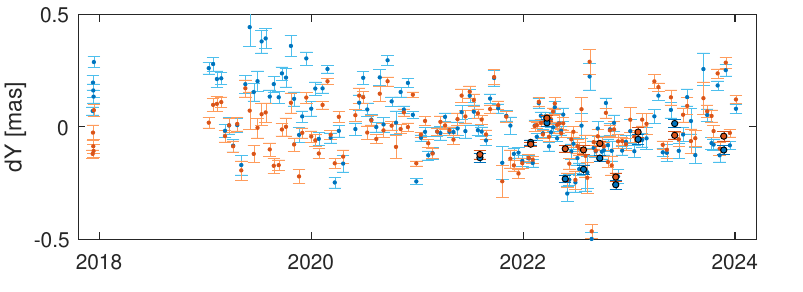}
      \caption{Nutation time series (dX, dY) with respect to IERS 20C04 from solution S1 (blue, fixed ICRF3-SX) and S2 (orange, fixed \crfname{}). The estimates from VGOS-R\&D are plotted as larger circles with black edges.}
         \label{dXdY}
\end{figure}

The median of the absolute deviation for celestial pole offsets is 70--80~\textmu as for both S1 and S2 with respect to IERS 20C04 with the exception of the dY between S1 and IERS 20C04. In Fig.~\ref{dXdY} we plot the time series of dX and dY from S1 (blue) and S2 (orange) with respect to the IERS 20C04 celestial pole offsets. For clarity we highlight the ten estimates coming from the VGOS-R\&D sessions with black edges.
It shows that there is a trend in the dY estimated from S1 before the year 2021, where dY from the solution with fixed \crfname{} are closer to IERS 20C04 (weighted mean: $84~/~3$~\textmu as, wrms: $175~/~116$~\textmu as, trend: $-48~/~21$~\textmu as/y computed from S1~/~S2 time series until 2021). The reason for this systematic difference, which disappears after 2021, is unclear to us at this time.

The statistics computed for the ten VGOS-R\&D sessions only, reveals wrms for the five EOP ($\mathrm{x_p}$, $\mathrm{y_p}$,  UT1-UTC, dX, and dY) from solutions S1~/~S2 with respect to IERS 20C04: 
$139~/~120$~\textmu as,
$109~/~128$~\textmu as,
$11~/~10$~\textmu s,
$40~/~56$~\textmu as, and
$86~/~60$~\textmu as.
Hence, one sees even from this limited dataset of ten sessions and regardless of the different scheduling techniques used in these sessions that there is an improvement of the geodetic performance of the  VGOS-R\&D compared with VGOS-OPS measured by the wrms of the detrended EOP series.

\begin{table*}
\caption{Statistics on EOP estimated from 155 VGOS sessions with respect to IERS 20C04. Values are provided for solution S1 (ICRF3-SX) and S2 (\crfname{}). }
\newcolumntype{Y}{>{\raggedleft\arraybackslash}X}
\label{tab:EOPstatVO}       
\begin{tabularx}{\textwidth}{l *{10}{Y} }
\hline
\hline
  & \multicolumn{2}{c}{~~~~~~$\mathrm{x_p}$  [\textmu as]} & 
  \multicolumn{2}{c}{~~~~~~$\mathrm{y_p}$ [\textmu as]} & 
  \multicolumn{2}{c}{~~~~~~UT1-UTC [\textmu s]} &
  \multicolumn{2}{c}{~~~~~~dX [\textmu as]} & 
  \multicolumn{2}{c}{~~~~~~dY [\textmu as]} \\
  & S1 & S2 & S1 & S2 & S1 & S2 & S1 & S2& S1 & S2 \\
\hline\noalign{\smallskip}
median                   & 33 & -28  &   6&-45    & -9&-8     & 9&5 & -7&-19    \\
weighted mean            & 56&5   &   7&-7     &   -10&-9 &   6&7 &  -9&-16\\
wrms$^{(*)}$               & 329&293  &   263&237  &   14&13 &     105&97  &   123&105\\
mean formal error        & 125&120  &   106&102  & 3&3     &   30&29 &   29&28\\
median of abs. deviation & 223&196  &   181&207 &   10&9   &   73&76   & 110&72\\
slope [per year]         & 40&61    &   -39&-25   &   -1&-1   &   2&-15   &   -45&0\\
\noalign{\smallskip}\hline
\end{tabularx}
\tablefoot{ $^{(*)}$ The wrms is computed for the detrended EOP time series after subtracting the slope.}
\end{table*}

\section{Conclusions}
\label{sec_conc}
In this paper we introduced the \crfname{} comprised of positions of 418 radio sources computed from $1.39\cdot10^6$ broadband group delays, available at \href{https://doi.org/10.48436/m6j36-s1082}{https://doi.org/10.48436/m6j36-s1082} \citep{Krasna2024}. In the global solution, we analyzed 145 VGOS-OPS and ten VGOS-R\&D sessions until December 2023. Our conclusions are as follows:\\
\begin{enumerate}
\item The \crfname{} has excellent source position precision. The median precision is 30~\textmu as in $\alpha^*$ and 47~\textmu as in $\delta$ computed over all 418 radio sources (formal errors directly from the least-squares adjustment).
\item The ten VGOS-R\&D sessions play a significant role in extending the VGOS source list. Several adjustments of the new source-centric scheduling strategy enabled the addition of 172 radio sources from only four VGOS-R\&D sessions with the potential to expand  the number of sources even further. The source-centric scheduling strategy \citep{Schartner2023b}, fully utilized in VR2301 and VR2302, is beneficial for generating a CRF since it allows one to observe more sources with a better distribution of scans among them within a 24-hour session.
\item The systematic distortions of \crfname{} versus ICRF3-SX for the common 408 radio sources (ten outliers eliminated) in terms of the VSH up to the degree and order two, reach in absolute values about 60~\textmu as for $R_1$, $D_2$, and $a_{2,1}^{e,Re}$. We note that removing 85 outliers ($D_n > 5$) reduces the VHS distortions by about a factor of two, suggesting that source structure in these outlier sources might be a significant contributor to distortions in the current VGOS network.
\item In \crfname{} in addition to the NNR condition, we stabilized the solution by constraining the slope in declination estimates versus declination to agree with ICRF3-SX. This was necessary because of  the lack of good observations from the southern hemisphere in this  dataset. Therefore, in order to limit VSH distortions in the CRF by the VGOS data itself, new southern stations will need to be added to the network, and the existing southern stations maintained. We especially commend the efforts at Hartebeesthoek, South Africa, and La Plata, Argentina, to join the VGOS network and hope they will be part of the network soon.
\item We identified ten radio sources as outliers with a normalized angular separation $D_n$ with respect to ICRF3-SX greater than 15. We provided a detailed examination of the position time series of the four sources with the largest normalized angular separation with respect to ICRF3-SX:  0119+115, 0229+131, 0355+508 (NRAO150), and 0642+449. By lowering the limit of $D_n$ to 5, the number of outliers becomes 85.\\
Recently, a study done by \citet{Petrov2024} concluded that frequency-dependent source position errors do not exceed 70~\textmu as and that the development of frequency-dependent reference frames of the entire sky is therefore not warranted for applications requiring accuracy at crude levels.\\
We show that the removal of S/X data prior to 2019.0 reduces the offset between the position from the S/X and VGOS data by several tens of percent for the discussed outlier sources (Tables~\ref{tab:souout} and \ref{tab:souwstd}). Common adjustment of all S/X and VGOS sessions would unnecessarily degrade the accuracy of the source position that is achieved with VGOS-only adjustment, especially for sources with observed variable positions over time as shown in Sect.~\ref{sec_compare_icrf3}. Therefore, we do not recommend computing a single CRF from all available VLBI sessions observed at different radio frequencies for applications requiring the highest accuracy. Our conclusion is supported by evaluating \crfname{} in the geodetic analysis of VGOS sessions. 
\item  We show that fixing to the \crfname{} instead of the ICRF3-SX a priori in the single-session analysis improves the wrms of the baseline length up to 3~mm, especially for the extremely long baselines (>\num{12 000}~km). An improvement larger than 1~mm is achieved on the majority of baselines longer than 8000 km. 
Additionally, a difference up to 2~mm in terms of a weighted mean between the estimated baseline length time series in these two solutions is obtained for several long baselines.
This difference in the baseline length offset and repeatability is essential for meeting the strategic goals of 1~mm station position accuracy and 0.1~mm/yr velocity stability set by the GGOS of the IAG.
\item In this paper we also studied the estimation of the path delay in the neutral atmosphere. In the \crfname{} we estimated zenith wet delay every 30 min and horizontal gradients every three hours. We show that in the current VGOS network, higher estimation rates for zwd (10 or 5 min) lower the mean formal errors for $\delta$ estimates for sources with declinations between approximately $-30\degr$ and $20\degr$ by about 12~\textmu as. On the other hand, there is an increase in  the position formal errors for southern sources with declination lower than approximately $-30\degr$ caused by the poor sky coverage of the isolated southern telescopes. Therefore, to tap the full potential of the VGOS also for the southern telescopes, construction of new VGOS telescopes serving as observing partners in the southern hemisphere is of the utmost importance.
\item In the EOP study, we concentrated on the effect of fixing \crfname{} instead of ICRF3-SX a priori on the estimated EOP. 
Our results show that the fixed \crfname{} instead of the fixed ICRF3-SX lowers the wrms of all five EOP with respect to IERS 20C04 of about 10--15\%. 
Furthermore, the statistics reveal that the wrms computed over VGOS-OPS (145 sessions) versus VGOS-R\&D (ten sessions) is considerably lower for the VGOS-R\&D sessions, although caution is advised due to the limited number of VGOS-R\&D sessions. The discussed wrms values by fixed \crfname{} are
293 versus 120~\textmu as for x-pole,
237 versus 128~\textmu as for y-pole,
13 versus 10~\textmu s for UT1-UTC,
97 versus 56~\textmu as for dX, and
105 versus 60~\textmu as for dY.
Based on this evidence of superior geodetic performance of the VGOS-R\&D sessions, we strongly support making the source-centric mode the nominal default scheduling strategy for VGOS sessions.
\end{enumerate}

\begin{acknowledgements}
The authors thank Zinovy Malkin, the referee, for his comments and suggestions which helped to improve the manuscript. The authors acknowledge the International VLBI Service for Geodesy and Astrometry (IVS) and all its components for providing VLBI data. HK and PC acknowledge the Austrian Science Fund (FWF) [10.55776/P35920]. The research was carried out in part at the Jet Propulsion Laboratory, California Institute of Technology, under a contract with the National Aeronautics and Space Administration (80NM0018D0004). Copyright 2024. All rights reserved.
\end{acknowledgements}

%
%
\bibliographystyle{aa} 
\bibliography{reference_krasna} 

\end{document}